\documentclass{elsart}

\usepackage{amsmath}
\input{epsf}
\begin{document}

\newcommand{\ra}{\rightarrow}
\newcommand{\pd}[2]{\frac{\partial {#1}}{\partial {#2}}}

\begin{frontmatter}
\title{Derivation of the probability distribution function for the
 local density of states of a disordered quantum wire via the 
replica trick and supersymmetry}

\author{J. E. Bunder and Ross H. McKenzie\thanksref{email}}

\address{School of Physics, University of New
South Wales, Sydney 2052, Australia}

\thanks[email]{New address: Department of Physics,
University of Queensland, Brisbane, 4072, Australia;
e-mail: mckenzie@physics.uq.edu.au}

\begin{abstract}
We consider the statistical properties of the
local density of states of a one-di\-men\-sional Dirac
equation in the presence of various types of disorder
with Gaussian white-noise distribution.
It is shown how either the
replica trick or supersymmetry can be used to calculate
exactly all
the moments of the local density of states.
Careful attention is paid to how the results change if the local
density of states is averaged over atomic length scales.
For both the replica trick and supersymmetry
the problem is reduced to finding the ground state
of a zero-dimensional Hamiltonian which is
written solely in terms of a pair of coupled ``spins''
which are elements of $u(1,1)$.
This ground state is explicitly found for the
particular case of the Dirac equation corresponding
to an infinite metallic quantum wire with a single
conduction channel.
The calculated moments of the local density of
states agree with those found previously
by Al'tshuler and Prigodin
[Sov. Phys. JETP 68 (1989) 198]
using a technique based on recursion relations
for Feynman diagrams.

\end{abstract}


\end{frontmatter}


\section{Introduction}
In the presence of disorder
mesoscopic systems such as quantum wires
and quantum dots have 
electronic properties which are characterised
by large statistical fluctuations between
different samples~\cite{imry}. For example, the conductance
of a typical sample can be quite different
from the ensemble averaged conductance.
A widely studied phenomena,
both experimentally and theoretically, is that
of universal conductance fluctuations:
the root mean square of the conductance
is of order $e^2/h$ (where $e$ is the electronic
charge and $h$ is Planck's constant) and is independent
of the strength of the disorder or the sample size~\cite{been-rev}.

Field theoretical techniques
based on supersymmetry and the replica trick
have proven to be extremely useful in 
studying non-perturbative effects associated
with disorder in non-interacting electron 
systems.
Supersymmetry has been used to evaluate
exactly the disorder averaged density of states
for a one-dimensional Schr{\" o}dinger equation
with a random white-noise potential~\cite{bohr},
various random one-dimensional Dirac
equations~\cite{hayn,hayn2,balents,bocquet},
and a number of models
associated with the lowest Landau level of a two-dimensional
electron gas in a high magnetic field~\cite{landau}.
The replica trick has been used to evaluate
exactly the disorder averaged density of states
of the same one-dimensional models involving the
Schr{\" o}dinger equation~\cite{tao},
and  Dirac equation~\cite{bouch2,bocquet}.
Non-linear sigma models based on either the replica
trick~\cite{wegner}
 or supersymmetry~\cite{efetov-book,zirn86,mirlin1,mirlin} have 
also proven to
be extremely powerful tools for studying Anderson
localisation in non-interacting electron systems.
Supersymmetric spin chains describe the delocalisation transition
associated with the quantum Hall effect~\cite{marston}. 


In a disordered electronic system static
fluctuations in the local electric potential
will produce statistical fluctuations in
the local density of states.
The corresponding probability
distribution function 
was first studied by Al'tshuler, Kravtsov,
and Lerner for the case of two and $2+\epsilon$
dimensions~\cite{akl}. Their results were based on a perturbative
renormalisation group analysis of
the non-linear sigma model introduced by Wegner~\cite{wegner}.
They found that near the mean value of the 
density of states the distribution function
was Gaussian but slowly decayed with tails of
logarithmically normal form.    
Near the metal-insulator transition due to Anderson localisation
the distribution approached a logarithmically normal form
for all values of the local density of states.
Mirlin and Fyodorov~\cite{mirlin1,mirlin} have developed a
 supersymmetric formalism to calculate all the moments of the
local density of states for a graded non-linear sigma
model relevant to disordered   systems in one, two, and three
dimensions  and a quasi-one-dimensional wire.
A novel path-integral approach was used by
Kolokolov to calculate the probability
distribution function for the inverse
participation ratio of both a finite and
an infinite disordered one-dimensional system~\cite{kolokolov}.

Al'tshuler and Prigodin~\cite{altshuler} evaluated
exactly all of the moments
of the local density of states for a one-dimensional metal.
They applied the
Berezinskii diagrammatic technique~\cite{Ber}
to the microscopic Hamiltonian.
From the moments it is straight-forward to
construct the complete probability distribution function.
They also showed that due to the
resulting distribution in Knight shifts
the  distribution function could be related 
to the line shape of a nuclear magnetic resonance (NMR) line.
With decreasing temperature and increasing disorder
the line shape becomes increasingly asymmetric,
with large differences between the average and typical values.
Recently, the results of Reference~\cite{altshuler}
were generalized to the case of a finite-sized ring
threaded by a magnetic flux~\cite{feldmann}.
Efetov and Prigodin have used the supersymmetry method
to calculate the NMR line shape   
for a small metallic particle which corresponds
to the case of zero dimensions~\cite{efpr,efetov2}.

Due to advances in nanotechnology it is
now possible to experimentally test some
of the theoretical predictions concerning
fluctuations in the local density of states.
Recent nuclear magnetic resonance experiments on monodisperse 
molecular
metal clusters of platinum atoms have tested the
theoretical predictions for zero-dimensions
and find a broad asymmetric line but that the density
of states fluctuations are obscured by additional
contributions from surface effects~\cite{fritschij}.
Experiments have imaged the local 
density of states in a three-dimensional disordered
metal (heavily doped GaAs)
by measuring the differential
conductance associated with resonant tunneling
through impurity states in asymmetric double-barrier
heterostructures~\cite{schmidt}. Fluctuations in the local
density of states were detected and found
to be enhanced in a magnetic field~\cite{holder}.

In this paper we show in detail how
both the supersymmetry and replica trick
methods can be used to derive all the moments of
local density of states for a random
one-dimensional Dirac equation.
Balents and Fisher~\cite{balents} recently used
supersymmetry to evaluate the disorder-average
of the one-particle Green function
for a one-dimensional Dirac equation
with a random mass.  Bouquet~\cite{bocquet}
used group theory to simplify their analysis
and also treated the problem using the replica trick.
Our analysis extends their  formalism
to allow for the evaluation of the
disorder average of products of Greens functions.
This is explicitly evaluated for the case of a quantum wire.
Our results agree with those of Reference~\cite{altshuler}.
However, in our view the field theoretic method
used here is more transparent and less cumbersome
than the Berezinskii technique.

In section \ref{sec-model} we introduce the model
Hamiltonian, the one-dimensional 
random Dirac equation and briefly mention
the different physical systems it is relevant
to, ranging from quantum wires to random
spin chains. In section \ref{sec-prob} we introduce
different forms of the local density of states and
define the generalized
participation ratios that are a measure of the
spatial extent of localised wave functions.
 We show how the
the moments of the density of states and the
participation ratios are related to one another.
 These quantities can be expressed in
terms of the disorder average of products
of retarded and advanced  Green's functions.
 Section \ref{sec-path} shows how these products of Green's functions
may be written in path integral form. 
Either the replica trick or supersymmetry can then
be used to evaluate exactly the average over disorder.
One is then left with a one-dimensional interacting
field theory.
 Section \ref{sec-trans} shows how a transfer matrix approach
can be used to reduce this to studying a
zero-dimensional  transfer Hamiltonian.
For both the replica trick and supersymmetry this Hamiltonian 
involves a pair of ``spin'' operators from  the non-compact 
algebra $su(1,1)$.
Section \ref{sec-ground} shows how
the moments of the local density of states can
be written in terms of the expectation value of
components of  these operators in the ground state of
the transfer Hamiltonian.
 In section \ref{sec-randommass} we find the
ground state for the particular case
of the random 
Dirac equation corresponding to a quantum wire.
This is then used to evaluate all the
moments of the density of states. In section \ref{sec-fulldos} 
we make some corrections to the moments of the density of states. 
In section \ref{sec-dist} we use the moments of the density of states
to calculate the distribution of the density of states.       

\section{The model}
\label{sec-model}
The one dimensional random Dirac equation Hamiltonian, $h$, for a 
relativistic particle with mass $M$ in a magnetic field $\Phi$ 
and in a scalar 
potential $V$ is a $2\times 2$ matrix
\begin{equation}
h=-i\sigma_z\partial_x+\Phi(x)\sigma_x+M(x)\sigma_y+V(x).\label{eq:h}
\end{equation}
The $\sigma_j$'s are Pauli spin matrices. For constant $\Phi$, $M$ and
$V$ this model describes a one 
dimensional semiconductor when particles are near the Fermi energy      
with 
a narrow forbidden band. The Hamiltonian is a $2\times 2$ matrix 
because it describes particles moving in both directions in one
dimension. The wave function is of the form $\Psi=(\psi_{\uparrow},
\psi_{\downarrow})$ where $\psi_{\uparrow}$ and $\psi_{\downarrow}$
correspond to right and left moving electrons, respectively.

The variables $\Phi$, $M$ and $V$ are random variables with a one 
dimensional spatial dependence. We can separate each of these random 
variables into an average part and a random part
\begin{equation}
\Phi(x)=\phi+{\tilde \phi}(x),\qquad M(x)=m+{\tilde m}(x), 
\qquad V(x)=v+{\tilde v}(x).
\end{equation}
The random part of each of the three variables is assumed to have a 
Gaussian white noise distribution centred about zero. If $D_{\phi}$ is 
the disorder strength of ${\tilde \phi}(x)$
\begin{equation}
\langle{\tilde \phi}(x){\tilde \phi}(x')\rangle=2D_{\phi}\delta(x-x'),
\qquad \langle{\tilde \phi}(x)\rangle=0,
\end{equation}
and similarly for ${\tilde m}(x)$ and ${\tilde v}(x)$. The disorder
strength of ${\tilde m}(x)$ and ${\tilde v}(x)$ is $D_m$ and $D_v$
respectively. We shall use the general form of the Dirac equation 
shown in equation (\ref{eq:h}) up until section \ref{sec-randommass}.

The random mass Dirac equation in a random magnetic field is when
there is no scalar potential, $V=0$. It also known as an 
incommensurate Dirac equation. The Green's function and localised 
density of states and has been calculated for this model
~\cite{golub,abrik}. The incommensurate 
Dirac equation is equivalent to the random $XX$ spin chain in a random 
magnetic field~\cite{mckenzie,bunder}.  In~\cite{balents} Balents and 
Fisher consider a random mass Dirac equation where $\Phi=V=0$ so only 
$M$ is non-zero. This is a commensurate Dirac 
equation~\cite{ovchin,hayn}. The commensurate 
Dirac equation is equivalent to the $XY$ spin chain in 
a random transverse magnetic~\cite{mckenzie,bunder}. 
Bocquet~\cite{bocquet} considered a number 
different cases, the random mass model (like~\cite{balents}), the 
massless particle in a scalar potential ($M=\Phi=0$) and a multiple 
disorder model ($m=\phi=v$).

In section \ref{sec-randommass} we shall consider the special case of 
the incommensurate Dirac equation. The average parts of $M$ and $\Phi$ 
vanish, $m=\phi=0$, and there is no scalar potential, $V=0$. We make 
the further assumption that the disorder strength of the mass and the 
magnetic field are equal, $D_{\phi}=D_m$. This corresponds to a
disordered quantum wire with a single channel.

\section{The probability distribution function}
\label{sec-prob}
\subsection{The different density of states}
The local density of states at the position $x$ for a general 
one-dimensional system with Hamiltonian $\bar{h}$
is defined by
\begin{equation}
\rho(E,x)=\langle x|\delta(E-\bar{h})|x
\rangle_Q=\sum_j|\psi_j(x)|^2\delta(E-e_j),\label{eq:density}
\end{equation}
with eigenfunction $\psi_j$ and eigenvalue $e_j$.
The subscript $Q$ indicates a quantum average (which uses the
eigenfunctions as weights) rather than a disorder
average.
For a finite closed sample the electron states make up a discrete
spectrum. However, the function $\rho$ can only take on two values,
zero or infinity. We can define a new and more appropriate density 
of states from~\cite{altshuler}
\begin{equation}
\rho_f(E,x)=\int_{-\infty}^{\infty}\rho(E',x)f(E-E')dE'
\label{eq:gend},
\end{equation} 
where $f$ is some arbitrary weight. We define the density of
states $\rho_1(\eta,E,x)$ as the density derived from the
weight 
\begin{equation}
f_{\eta}(x)=\frac{1}{\pi}\frac{\eta}{\eta^2+x^2},
\label{eq:delta}
\end{equation}
where $\eta$ is small, positive and real. This weight is 
equivalent to the Dirac delta function, $\delta(x)$,
in the limit $\eta\ra 0$. To obtain the thermodynamic limit of 
$\rho_1,$
 $L\ra\infty$, we take the limit $\eta\ra 0$. In the thermodynamic
limit  $\rho(E,x)=\rho_1(0,E,x)$.

A full one-dimensional wave function of the general one-dimensional
system for states near the Fermi energy
is related to the spinor 
$\Psi=(\psi_{\uparrow},\psi_{\downarrow})$ in the Dirac equation by
\begin{equation}
\psi(x)=\psi_{\uparrow}(x)e^{ik_Fx}+\psi_{\downarrow}(x)e^{-ik_Fx},
\end{equation}
where $\psi_{\uparrow}$ and $\psi_{\downarrow}$ denote the amplitudes
for right and left moving electrons, respectively. These amplitudes
are slowly varying on atomic length scales of order $k_F^{-1}$ but the
exponential terms vary rapidly over this length scale. The
intensity of the total wave function is
\begin{equation}
|\psi(x)|^2=|\psi_{\uparrow}(x)|^2+|\psi_{\downarrow}(x)|^2+
\psi_{\uparrow}^*(x)\psi_{\downarrow}(x)e^{-2ik_Fx}+
\psi_{\downarrow}^*\psi_{\uparrow}(x)e^{2ik_Fx}.\label{eq:rapid}
\end{equation}
Thus, the local density of states defined in equation
(\ref{eq:density}) contains terms which oscillate rapidly over atomic
length scales. We can define an alternative density of states that
does not include these rapid oscillations
\begin{equation}
{\tilde \rho}(E,x)=\sum_{j}\left[|\psi_{j\uparrow}(x)|^2
+|\psi_{j\downarrow}(x)|^2\right]\delta(E-e_j).\label{eq:density2}
\end{equation}

We can write an equation similar to (\ref{eq:gend}) but with ${\tilde
\rho}$ substituted for $\rho$ 
\begin{equation}
{\tilde \rho}_f(E,x)=\int_{-\infty}^{\infty}{\tilde \rho}(E',x)
f(E-E')dE'\label{eq:gend2}.
\end{equation} 
We define ${\tilde \rho}_1(\eta,E,x)$ as we defined $\rho_1(\eta,E,x)$,
using the weight $f_{\eta}$ in equation (\ref{eq:delta}), but the 
former is obtained from equation (\ref{eq:gend2}) rather than
equation (\ref{eq:gend}). We shall find, as did Al'tshuler and
Prigodin~\cite{altshuler}, that these two different local density of
states have quite different probability distribution functions.  

For the Dirac equation the Green's function is a $2\times 2$ matrix. 
A component of the retarded and advanced Green's function matrix
 is defined as
\begin{equation}
G^{R,A}_{\alpha,\beta}(E,x,x)=\left\langle x,\alpha\left|\frac{1}{E\pm
i\eta-h}\right|x,\beta\right\rangle_Q,\label{eq:green}
\end{equation}
where $\alpha$ and $\beta$ are spin indices.
In this paper we are interested in the trace
 of this matrix, $G^{R,A}(E,x,x)=\mathrm{Tr_{spin}}
G_{\alpha,\beta}^{R,A}(E,x,x)$. We shall refer to $G^{R,A}$ as the 
retarded  or advanced Green's function. In terms of Green's functions 
the density of states is
\begin{equation}
{\tilde \rho}_1(\eta,E,x)=\frac{1}{2i\pi}\left(G^A(E,x,x)-
G^R(E,x,x)\right).\label{eq:greendos}
\end{equation}

\subsection{Participation ratio}
There are a number of different criteria to distinguish energy regions 
of extended and localized states of a particle in a random potential. 
The inverse participation ratio, $P_j$, is the sum over all positions 
of the fourth power of the $j$th eigenfunction
\begin{equation}
P_j=\sum_x|\psi_j(x)|^4.
\end{equation}
The eigenfunction at position $x$ is $\psi_j(x)$ with a 
corresponding eigenvalue $e_j$ and Hamiltonian $\bar{h}$ 
which acts on the wave function $|x\rangle$ as follows
\begin{equation}
\bar{h}|x\rangle=\bar{h}\sum_j\psi_j(x)=\sum_je_j\psi_j(x).
\end{equation}
The inverse participation ratio is positive for localized states but 
vanishes for extended states in the thermodynamic limit. The 
participation ratio averaged over the ensemble (i.e. the random 
potential), $\langle{P_j}\rangle$, measures those sites which make 
a significant contribution to the eigenfunction normalization. The
participation ratio is related to how much the wave functions extend
through space. For localized states 
$\langle{P_j}\rangle$ is nonzero. If the eigenstates 
are completely delocalized then 
$\langle{P_j}\rangle$ vanishes for an infinite system. 

Instead of considering $P_j$ which is for individual eigenstates of 
individual systems we consider the ensemble averaged quantity in a 
narrow energy window (spectral average) centred about the energy $E$
\begin{equation}
P^{(2)}(E)=\left\langle\sum_{j,x}|\psi_j(x)|^4\delta(E-e_j)\right
\rangle\left\langle\sum_j\delta(E-e_j)\right
\rangle^{-1},\label{eq:ptemp}
\end{equation}
where the angled brackets refer to the average over the ensemble, 
or the disorder average. Once we take the disorder average of some
quantity it will no longer depend on the position $x$. 
Hence, after summing equation (\ref{eq:density}) over all $x$ 
and taking the disorder average we obtain
\begin{equation}
2L\langle\rho(E)\rangle=\sum_j\delta(E-e_j),
\end{equation}
where $2L$ is the length of the system.
Similarly, the average in the numerator of equation (\ref{eq:ptemp}) 
does not depend on the position, $x$, so
\begin{equation}
P^{(2)}(E)\langle\rho(E)\rangle=\left\langle\sum_j|\psi_j(x)|^4
\delta(E-e_j)\right\rangle.\label{eq:p2}
\end{equation}

We can generalise equation (\ref{eq:p2}) to
\begin{equation}
P^{(k)}(E)\langle\rho(E)\rangle=\left\langle\sum_j|\psi_j(x)|^{2k}
\delta(E-e_j)\right\rangle,\label{eq:pratio}
\end{equation}
where $P^{(k)}$ is the $k$th moment of the participation ratio. 
The first moment ($k=1$) of the participation ratio is unity.

\subsection{Moments of the local density of states}

The $k$th moment of the local density of states, $\rho_1$, is the disorder
average of the $k$th power of the density of states,
$\langle\rho_1(\eta,E)^k\rangle$.  
We can write the $k$th moment of the density of states
 in terms of the density of states, $\rho$,
\begin{multline}
\langle\rho_1(\eta,E)^k\rangle=\int_{-\infty}^{\infty} dE_1\ldots 
dE_kf_{\eta}(E-E_1)\ldots f_{\eta}(E-E_k)\\
\times\langle\rho(E_1)\rho(E_2)\ldots\rho(E_k)\rangle.
\end{multline}
Wegner~\cite{wegner} showed that in the limit $\eta\ra 0$ 
\begin{equation}
\langle\rho_1(\eta,E)^k\rangle=\left\langle\sum_j|\psi_j|^{2k}
\delta(E-e_j)\right\rangle
\int_{-\infty}^{\infty}dE_1f_{\eta}(E-E_1)^k.
\end{equation}
The disorder average part is the $k$th moment of the  
participation ratio defined in 
equation (\ref{eq:pratio}) so the $k$th moment of the density of
states is related to the $k$th moment of the participation 
ratio by
\begin{equation}
\langle\rho_1(\eta,E)^k\rangle=P^{(k)}\langle\rho(E)\rangle
\int_{-\infty}^{\infty}dE_1f_{\eta}(E-E_1)^k.\label{eq:rho_1}
\end{equation}
We can evaluate the integral over the weight function
\begin{equation}
\int_{-\infty}^{\infty}dxf_{\eta}(x)^k=(4\pi\eta)^{1-k}
\frac{\Gamma(2k-1)}{\Gamma(k)^2},
\end{equation}
so that
\begin{equation}
\langle\rho_1(\eta,E)^k\rangle=(4\pi\eta)^{1-k}
\frac{\Gamma(2k-1)}{\Gamma(k)^2}P^{(k)}\langle\rho(E)\rangle
.\label{eq:mom-pratio}
\end{equation}
From equation (\ref{eq:pratio}) $P^{(1)}=1$, since the left hand side
is $\langle\rho(E)\rangle$. This agrees with the above equation in the
thermodynamic limit since in this limit $\rho_1=\rho$. 

The general equation (\ref{eq:gend}) for the density of states
$\rho_f$ can be generalised to resemble equation (\ref{eq:rho_1})
~\cite{altshuler}
\begin{equation}
\langle\rho_f(E)^k\rangle=P^{(k)}\langle\rho(E)\rangle
\int_{-\infty}^{\infty}dE_1f(E-E_1)^k.\label{eq:rho_f}
\end{equation}
The moments of the participation ratio are independent of the
weight $f$. We can write a similar equation for the moments of
${\tilde\rho}_f$ by replacing $\rho$ with ${\tilde\rho}$ in the above
equation. The $k$th moment of ${\tilde \rho}_1$ can be written in 
terms of Green's functions by using equation (\ref{eq:greendos})
\begin{equation}
\langle{\tilde\rho}_1(\eta,E)^k\rangle=\frac{k\mathrm{!}}
{(2i\pi)^k}\sum_{j=0}^k\frac{(-1)^j}{j\mathrm{!}(k-j)\mathrm{!}}
\left\langle G^A(E,x,x)^{k-j}
G^R(E,x,x)^j\right\rangle.\label{eq:mom}
\end{equation}
In this paper we calculate the Green's function correlator
\begin{equation}
\left\langle G^A(E,x,x)^{k-j}G^R(E,x,x)^j\right\rangle. 
\end{equation}
This correlator enable us to calculate the moments of
the density of states, $\langle{\tilde\rho}_1^k\rangle$. By extending
this method a little we can also calculate $\langle\rho_1^k\rangle$. 
 This allows us to calculate the moments of the 
participation ratio via equation (\ref{eq:mom-pratio}).

In section \ref{sec-randommass} we will calculate the moments of 
the local density of states, $\langle{\tilde
\rho}_1(\eta,E)^k\rangle$, for a random mass Dirac equation. 
In section \ref{sec-fulldos} we
find $\langle\rho_1(\eta,E)^k\rangle$ for the same model  
which gives the moments of the participation ratio, $P^{(k)}$. 
Knowing the moments of the participation
ratio we can solve equation (\ref{eq:rho_f}) for
$\langle\rho_f^k\rangle$ using a general weight, 
$f$. In section \ref{sec-dist} we take $f$ to be the derivative 
of the Fermi-Dirac distribution function.  

\section{Path integral form of products of Green's functions}
\label{sec-path}
A component of the advanced and retarded Green's function matrix 
 at some position $x$ and energy $E$ is defined in equation 
(\ref{eq:green}). 
Note that the retarded Green's function is the complex 
conjugate of the advanced Green's function.
A solution for the Green's function can often be obtained by writing
it in terms of a path integral. There are two standard ways to
construct the path integral, the replica trick and supersymmetry. 
Both techniques allow the average over the disorder to be performed
analytically.  

\subsection{Replica trick}
To construct the replica trick path integral form of the
 advanced Green's function we firstly write it as a path integral 
over two complex variable, $s$ and $s^*$,
\begin{multline}
G^A(E,x,x)=i\left[\int\mathcal{D}s^*
\mathcal{D}ss^*(x)s(x)\right.\\
\left.\times\exp\left(-i\int_{-L}^{L}dys^*(y)[E-i\eta-h]s(y)
\right)\right]\\
\times\left[\int\mathcal{D}s^*\mathcal{D}s\mathcal{D}\exp
\left(-i\int_{-L}^{L}dys^*(y)[E-i\eta-h]s(y)\right)\right]^{-1}
\end{multline}
The region of integration $(-L,L)$ is over the length of the system.
We shall later take the limit $L\ra\infty$.   
The retarded Green's function is the complex conjugate 
of the trace of the advanced Green's function
\begin{multline}
G^R(E,x,x)=-i\left[\int\mathcal{D}s^*
\mathcal{D}ss^*(x)s(x)\right.\\
\left.\times\exp\left(i\int_{-L}^{L}dys^*(y)[E+i\eta-h]s(y)
\right)\right]\\
\times\left[\int\mathcal{D}s^*\mathcal{D}s\mathcal{D}\exp
\left(i\int_{-L}^{L}dys^*(y)[E+i\eta-h]s(y)\right)\right]^{-1}.
\end{multline}
The dominator is $Z$, the partition function. 

If the variables $s$ and $s^*$ are replicated $n$ times the advanced
Green's function becomes
\begin{multline}
G^A(E,x,x)=\frac{i}{n}\left[
\int\mathcal{D}s^*\mathcal{D}s\sum_{l=1}^ns^*_l(x)s_l(x)
\right.\\
\left.\times\exp\left(-i\int_{-L}^{L}dy\sum_{l=1}^ns_l^*[E-i\eta-h]
s_l\right)\right]\\
\times\left[\int\mathcal{D}s^*\mathcal{D}s\exp
\left(-i\int_{-L}^{L}dy\sum_{l=1}^ns_l^*[E-i\eta-h]s_l\right)
\right]^{-1}.
\end{multline}
The dominator is now $Z^n$. Thus, if we let $n\ra 0$ the denominator 
is unity and
\begin{multline}
G^A(E,x,x)=\lim_{n\ra 0}\frac{i}{n}
\left[\int\mathcal{D}s^*\mathcal{D}s\sum_{l=1}^ns^*_l(x)
s_l(x)\right.\\
\left.\times\exp\left(-i\int_{-L}^{L}dy\sum_{l=1}^ns_l^*
[E-i\eta-h]s_l\right)\right].
\end{multline}
This is one form of the replica trick path integral. A similar
expression may be constructed for the retarded Green's function. There
are other forms of the replica trick path integral which only differ
in the placement of the complex number $i$, such as the form
used by Bocquet~\cite{bocquet}.

We shall now construct a replica trick form of a multiple of Green's
functions. The $j$th power of the advanced Green's function is
\begin{multline}
G^A(E,x,x)^j=\frac{i^j}{j\mathrm{!}}\\
\times\left[\int\mathcal{D}s^*\mathcal{D}s\left(s^*(x)s(x)
\right)^j\exp\left(-i\int_{-L}^Ldys^*(y)[E-i\eta-h]s(y)\right)\right]
\\
\times\left[\int\mathcal{D}s^*\mathcal{D}s\exp\left(
-i\int_{-L}^{L}dys^*(y)[E-i\eta-h]s(y)\right)\right]^{-1},
\end{multline}
and the $j$th multiple of the retarded Green's function is the complex
conjugate of this. 
The product of the $(k-j)$th power of the advanced Green's function
and the $j$th power of the retarded Green's function may be written as
 a path integral over four complex variables, $s$, $s^*$, $s'$
and $s^{\prime *}$, 
\begin{multline}
G^A(E,x,x)^{k-j}G^R(E,x,x)^j=\frac{i^k(-1)^j}{(k-j)
\mathrm{!}j\mathrm{!}}\\
\times\left[\int\mathcal{D}s^*\mathcal{D}s\mathcal{D}
s^{\prime *}\mathcal{D}s'(s^*(x)s(x))^{k-j}(s^{\prime *}(x)s'(x))^j
\right.\\
\left.\times\exp\left(-i\int_{-L}^{L}dy\left[s^*[E-i\eta-H]s
-s^{\prime *}[E+i\eta-h]s'\right]\right)\right]\\
\times\left[\int\mathcal{D}s^*\mathcal{D}s\mathcal{D}s^{\prime *}
\mathcal{D}s'\right.\\
\left.\times\exp\left(-i\int_{-L}^{L}dy\left[s^*[E-i\eta-h]s
-s^{\prime *}[E+i\eta-h]s'\right]\right)\right]^{-1}.
\end{multline}
If we replicate $s$ $n$ times and $s'$ $n'$ times then
\begin{multline}
G^A(E,x,x)^{k-j}G^R(E,x,x)^j=\frac{i^k(-1)^j
\Gamma(n)\Gamma(n')}{\Gamma(k-j+n)\Gamma(j+n')}\\
\times\left[\int\mathcal{D}s^*\mathcal{D}s\mathcal{D}
s^{\prime *}\mathcal{D}s'\left(\sum_{l=1}^ns^*_l(x)s_l(x)\right)
^{k-j}\left(\sum_{l=1}^{n'}s_l^{\prime *}(x)s'_l(x)\right)^j
\right.\\
\left.\times\exp\left(-i\int_{-L}^{L}dy\left[\sum_{l=1}^ns_l^*
[E-i\eta-h]s_l-\sum_{l=1}^{n'}s_l^{\prime *}[E+i\eta-h]s'_l\right]
\right)\right]\\
\times\left[\int\mathcal{D}s^*\mathcal{D}s\mathcal{D}s^{\prime *}
\mathcal{D}s'\right.\\
\left.\times\exp\left(-i\int_{-L}^{L}dy\left[\sum_{l=1}^n
s_l^*[E-i\eta-h]s_l-\sum_{l=1}^{n'}s_l^{\prime *}[E+i\eta-h]s'_l
\right]\right)\right]^{-1}.\label{eq:coeff}
\end{multline}
The prefactor $\Gamma(n)\Gamma(n')/\Gamma(k-j+n)\Gamma(j+n')$ is   
derived in appendix \ref{app-int}. When taking the limit
$n, n'\ra 0$, as before, the denominator is unity and
\begin{multline}
G^A(E,x,x)^{k-j}G^R(E,x,x)^j=\lim_{n,n'\ra 0}
\frac{i^k(-1)^j\Gamma(n)\Gamma(n')}{\Gamma(k-j+n)\Gamma(j+n')}\\
\times\int\mathcal{D}s^*\mathcal{D}s\mathcal{D}
s^{\prime *}\mathcal{D}s'\left(\sum_{l=1}^ns^*_l(x)s_l(x)
\right)^{k-j}\left(\sum_{l=1}^{n'}s_l^{\prime *}(x)s'_l(x)
\right)^j\\
\times\exp\left(-i\int_{-L}^{L}dy\left[\sum_{l=1}^n
s_l^*[E-i\eta-h]s_l-\sum_{l=1}^{n'}s_l^{\prime *}[E+i\eta-h]
s'_l\right]\right).\label{eq:replica}
\end{multline}

\subsection{Supersymmetry}

The supersymmetric path integral form of the
advanced Green's function is an integral over two complex 
variables, $s$ and $s^*$, and two
Grassman variables, $\xi$ and $\xi^*$~\cite{efetov,efetov-book},
\begin{multline}
G^A(E,x,x)=i\int\mathcal{D}\xi^*\mathcal{D}
\xi\mathcal{D}s^*\mathcal{D}ss^*(x)s(x)\\
\times\exp\left(-i\int_{-L}^{L}dy\left[\xi^*[E-i\eta-h]\xi+s^*
[E-i\eta-h]s\right]\right).
\end{multline}
Unlike the replica trick no denominator corresponding to a
normalization (or partition function) appears because the unwanted
contributions from the complex and Grassman variables cancel. The 
Grassman variables are introduced for this reason.
As stated before, the retarded Green's function is the complex
conjugate of the advanced Green's function. 
The $j$th multiple of the advanced Green's function is
\begin{multline}
G^A(E,x,x)^j=\frac{i^j}{j\mathrm{!}}\int\mathcal{D}
\xi^*\mathcal{D}\xi\mathcal{D}s^*\mathcal{D}s(s^*(x)s(x))^j\\
\times\exp\left(-i\int_{-L}^{L}dy\left[\xi^*[E-i\eta-h]\xi+s^*
[E-i\eta-h]s\right]\right).
\end{multline}

The product $G^A(E,x,x)^{k-j}G^R(E,x,x)^j$ can be 
written in terms of a supersymmetric path integral over eight
variables~\cite{efetov-book}
\begin{multline}
G^A(E,x,x)^{k-j}G^R(E,x,x)^j=\frac{i^k(-1)^j}{(k-j)
\mathrm{!}j\mathrm{!}}\\
\times\int\mathcal{D}\xi^*\mathcal{D}\xi\mathcal{D}s^*
\mathcal{D}s\mathcal{D}\xi^{\prime *}\mathcal{D}\xi'\mathcal{D}
s^{\prime *}\mathcal{D}s'(s^*(x)s(x))^{k-j}(s^{\prime *}(x)s'(x))^j\\
\times\exp\left(-i\int_{-L}^{L}dy\left[\xi^*[E-i\eta-h]\xi+s^*
[E-i\eta-h]s\right.\right.\\
\left.\left.-\xi^{\prime *}[E+i\eta-h]\xi'-s^{\prime *}[E+i\eta-h]s'
\right]\right).\label{eq:susy}
\end{multline}

\section{The transfer Hamiltonian}
\label{sec-trans}
We now briefly review the transfer matrix method
~\cite{feynman,bohr,hayn,hayn2} which can be used to reduce a 
one-dimensional
field theory into a zero-dimensional Schr{\" o}dinger-type equation.
We wish to calculate the disorder average of a product of Green's
functions, $\langle G^A(E,x,x)^{k-j}G^R(E,x,x)^j\rangle$. 
We have shown in the
previous section that this average is of the form
\begin{equation}
\langle\phi(x)\rangle_S=\left\langle\int\mathcal{D}X\phi(x) 
e^{-S(X)}\right\rangle,
\end{equation}
where $S$ is some action and $x$ is a spatial variable. For the
replica trick
\begin{equation}
\phi(x)=\left(\sum_{l=1}^ns^*_l(x)s_l(x)
\right)^{k-j}\left(\sum_{l=1}^{n'}s_l^{\prime *}(x)s'_l(x)
\right)^j,
\end{equation}
and for supersymmetry
\begin{equation}
\phi(x)=(s^*(x)s(x))^{k-j}(s^{\prime *}(x)s'(x))^j.
\end{equation}
The action is defined as the integral of the Lagrangian
\begin{equation}
S=\int_{-L}^L\mathcal{L}dy.
\end{equation}

All the disorder is contained in the Hamiltonian, $h$, which is
contained in the exponential term 
$e^{-S(X)}$. If we have a random Gaussian 
variable, $V$, which has an average value $v$ and disorder strength 
$D_v$ and $g$ is some non-random function then the average may be 
taken by using~\cite{hayn,hayn2}
\begin{equation}
\left\langle \exp \left(i\int_{-L}^Ldy V(y)g(y)\right)\right\rangle
=\exp\left(-D_v\int_{-L}^Ldyg(y)^2\right).\label{eq:ave}
\end{equation}
 
Once we have taken the disorder average we can find a Hamiltonian, 
known as the transfer Hamiltonian~\cite{bohr,hayn,hayn2}, from the 
disorder averaged Lagrangian.
If we have a coordinate $q$ and a momentum $p$ the transfer
Hamiltonian  is
\begin{equation}
H=\mathcal{L}-p\dot{q}\label{eq:trans}.
\end{equation}
A path integral is a sum over all possible configurations. We can
rewrite the path integral of $\langle\phi\rangle_S$ as such a sum. The
continuous interval $(-L,L)$ can be written as a regular lattice with
spacing $\alpha$ so that there are $2L/\alpha +1$ sites.
The value of the variable $X$ at site $x$ is
$X_x$. Because of periodic boundary conditions $X_{-L}=X_L$. The path
integral is now an integral over the variables $X_x$ with
$x=-L,-L+\alpha,\ldots, L$. Or more compactly
\begin{equation}
\langle\phi(x)\rangle_S=\int dX_LdX_x K(-L,x,X_L,X_x)\phi 
K(x,L,X_x,X_L),\label{eq:phi}
\end{equation}
where most of the integrals are hidden inside the functions $K$ which
may be thought of as propagators. Note that the $\phi$ inside the
integral longer has any
spatial dependence because we have already taken the disorder
average. 

From a propagator such as $K$ we can define a set of eigenfunctions, 
$\Psi_{mL}$ and $\Psi_{mR}$. The two eigenfunctions, $\Psi_{mL}$ and
$\Psi_{mR}$, are left and right eigenfunctions respectively. We need
to distinguish between the two because the transfer Hamiltonian need
not be Hermitian. The right eigenfunctions and propagator should 
satisfy an equation of the form
\begin{equation}
\Psi_{mR}(X_x)=\int dX_y K(y,x,X_y,X_x)\Psi_{mR}(X_y).
\end{equation}
Using Feynman's transfer matrix technique~\cite{feynman} we take 
$x=y+\alpha$ and 
find a Schr{\" o}dinger-type equation for $\Psi_{mR}$:
\begin{equation}
i\pd{\Psi_{mR}}{x}=H\Psi_{mR},
\end{equation}
where $H$ is the transfer Hamiltonian defined in equation 
(\ref{eq:trans}). The left eigenfunction satisfies a similar 
equation. We can define some
eigenstates, $E_m$, such that $H\Psi_{mR}=E_n\Psi_{mR}$ and 
$\Psi_{mL}H=E_n\Psi_{mL}$. We define 
the eigenfunctions to be orthonormal so that
\begin{equation}
\int dX\Psi_{mL}^*(X)\Psi_{nR}(X)=\delta_{mn}.
\end{equation} 

A propagator can be written in terms of its eigenfunctions and
eigenstates 
\begin{equation}
K(y,x,X_y,X_x)=\sum_m\Psi_{mR}(X_y)\Psi_{mL}^*(X_x)
\exp[-E_m(x-y)].
\end{equation}
We substitute this into equation (\ref{eq:phi}) and
 using the orthonormality of the eigenfunctions obtain
\begin{equation}
\langle\phi(x)\rangle_S=\sum_m\int|\Psi_m(X)|^2\phi dX 
 \exp[-2LE_m],
\end{equation}
where we have defined $|\Psi_m|^2=\Psi_{mL}^*\Psi_{mR}$.

We shall now derive an equation of the above form for the
Green's function correlator  
$\langle G^A(E,x,x)^{k-j}G^R(E,x,x)^j\rangle$. 
We obtain both a 
replica and a supersymmetric version. The equation can be further
simplified in the limit $L\ra\infty$.
 
\subsection{Replica trick}

Since the Hamiltonian, $h$, for the random Dirac equation (\ref{eq:h}) 
is a $2\times 2$ matrix the path integral variables in equation 
(\ref{eq:replica}) are two component vectors. For example
\begin{equation}
s=(s_{\uparrow},s_{\downarrow}).
\end{equation}
By substituting the Dirac Hamiltonian (\ref{eq:h}) into equation 
(\ref{eq:replica}) we can obtain the action  
\begin{multline}
S=-\int_{-L}^L dy\left[\sum_{l=1}^n\left[s_l^*\sigma_z\partial 
s_l+i\Phi s_l^*\sigma_xs_l+iMs_l^*\sigma_ys+(iV-\epsilon_-)s_l^*
s_l\right]\right.\\
\left.-\sum_{l=1}^{n'}\left[s_l^{\prime *}\sigma_z\partial s_l'+
i\Phi s_l^{\prime *}\sigma_xs_l'+iMs_l^{\prime *}\sigma_ys_l'+
(iV-\epsilon_+)s_l^{\prime *}s_l'\right]\right].
\end{multline}
The coordinates are
\begin{equation}
q_l=s_l=(s_{l\uparrow},s_{l\downarrow}), 
\qquad q_l'=s_l'=(s'_{l\uparrow},s'_{l\downarrow}),
\end{equation}
and so we can calculate the momenta
\begin{equation}
p_l=(s_{l\uparrow},s_{l\downarrow}), 
\qquad p_l'=(s'_{l\uparrow},s'_{l\downarrow}).
\end{equation}

To obtain the transfer Hamiltonian the action needs to be transformed
into the form of a coherent path integral~\cite{balents,bocquet}
\begin{alignat}{2}
s_{l\uparrow}&\ra b_{l\uparrow},&\qquad s_{l\downarrow}&\ra 
b_{l\downarrow}^{\dag},\nonumber\\
s_{l\uparrow}^*&\ra b_{l\uparrow}^{\dag},&\qquad s_{l\downarrow}^*
&\ra b_{l\downarrow},\label{eq:transform}
\end{alignat}
and $b_l=(b_{l\uparrow},b_{l\downarrow}^{\dag})$. The transformation 
of the primed terms is identical.
The coordinates and momenta in terms of these new variables are
\begin{alignat}{2}
q_l&=(b_{l\uparrow},b_{l\downarrow}^{\dag}),& \qquad p_l&=
(-b_{l\uparrow}^{\dag},b_{l\downarrow}),\nonumber\\
q_l'&=(b'_{l\uparrow},b_{l\downarrow}^{\prime\dag}),& 
\qquad p_l'&=(b_{l\uparrow}^{\prime\dag},-b'_{l\downarrow}).
\end{alignat}
The transformed action is
\begin{multline}
S=-\int_{-L}^L dy\left[\sum_{l=1}^n\left[b_{l\uparrow}^{\dag}\partial 
b_{l\uparrow}-b_{l\downarrow}\partial b_{l\uparrow}^{\dag}+
i\Phi(b_{l\uparrow}^{\dag}b_{l\downarrow}^{\dag}+
b_{l\downarrow}b_{l\uparrow})\right.\right.\\
\left.+M(b_{l\uparrow}^{\dag}b_{l\downarrow}^{\dag}-
b_{l\downarrow}b_{l\uparrow})+(iV-\epsilon_-)(b_{l\uparrow}^{\dag}
b_{l\uparrow}+b_{l\downarrow}b_{l\downarrow}^{\dag})\right]\\
-\sum_{l=1}^n\left[b_{l\uparrow}^{\prime\dag}\partial b'_{l\uparrow}
-b'_{l\downarrow}\partial b_{l\uparrow}^{\prime\dag}+
i\Phi(b_{l\uparrow}^{\prime\dag}b_{l\downarrow}^{\prime\dag}+
b'_{l\downarrow}b'_{l\uparrow})\right.\\
\left.\left.\left.+M(b_{l\uparrow}^{\prime\dag}b_{l\downarrow}
^{\prime\dag}-b'_{l\downarrow}b'_{l\uparrow})+(iV-\epsilon_+)
(b_{l\uparrow}^{\prime\dag}b'_{l\uparrow}+
b'_{l\downarrow}b_{l\downarrow}^{\prime\dag})\right]\right.\right].
\end{multline}

At this stage we take the ensemble average of the action using
equation (\ref{eq:ave}). 
After averaging over the disorder we can find the transfer
Hamiltonian~\cite{bohr,hayn,hayn2}. From equation (\ref{eq:trans}) 
the transfer Hamiltonian is
\begin{multline}
H=-2i\phi(Z_y-Z_y')+2im(Z_x-Z_x')-2i(v-E)(Z_z-Z_z')+2\eta(Z_z+Z_z')\\
+4D_{\phi}(Z_y-Z_y')^2+4D_m(Z_x-Z_x')^2+4D_v(Z_z-Z_z')^2,
\label{eq:ham}
\end{multline}
where 
\begin{alignat}{2}
{\bf Z}&=-\half \sum_{l=1}^np_l
(i\sigma_x,i\sigma_y,\sigma_z)q_l,
&\qquad {\bf Z}'&=\half \sum_{l=1}^np'_l(i\sigma_x,i\sigma_y,\sigma_z)
q'_l\nonumber\\
Z_x&={\textstyle{\frac{i}{2}}}\sum_{l=1}^n(b_{l\uparrow}^
{\dag}b_{l\downarrow}^{\dag}
-b_{l\downarrow}b_{l\uparrow}),&\qquad
Z'_x&={\textstyle{\frac{i}{2}}}\sum_{l=1}^{n'}
(b_{l\uparrow}^{\prime\dag}b_{l\downarrow}^{\prime\dag}-
b'_{l\downarrow}b'_{l\uparrow}),\nonumber\\
Z_y&=\half \sum_{l=1}^n(b_{l\uparrow}^{\dag}b_{l\downarrow}^{\dag}
+b_{l\downarrow}b_{l\uparrow}),&\qquad Z'_y&=
\half \sum_{l=1}^{n'}(b_{l\uparrow}^{\prime\dag}
b_{l\downarrow}^{\prime\dag}+b'_{l\downarrow}b'_{l\uparrow}),
\nonumber\\
Z_z&=\half \sum_{l=1}^n(b_{l\uparrow}^{\dag}b_{l\uparrow}
+b_{l\downarrow}b_{l\downarrow}^{\dag}),&\qquad Z'_z&=\half 
\sum_{l=1}^{n'}(b_{l\uparrow}^{\prime\dag}b'_{l\uparrow}
+b'_{l\downarrow}b_{l\downarrow}^{\prime\dag}).\label{eq:Zb}
\end{alignat}
Note that the transfer Hamiltonian does not depend on the spatial
coordinate $x$ making it a zero-dimensional equation.
We can define some ladder operators
\begin{equation}
Z_{\pm}=Z_y\mp iZ_x,
\end{equation}
and similarly for $Z_{\pm}'$.
The vector ${\bf Z}$ satisfies $su(1,1)$ commutation
rules~\cite{bocquet}
\begin{equation}
[Z_+,Z_-]=-2Z_z, \qquad [Z_z,Z_+]=Z_+, \qquad [Z_z,Z_-]=-Z_-,
\label{eq:comm}
\end{equation}
and also $(Z_+')^{\dag}=Z_-'$. Identical rules govern ${\bf Z}$.

We define the $m$th eigenstate of the transfer Hamiltonian as $E_m$ 
with corresponding eigenfunction $\Psi_m$. Since the transfer 
Hamiltonian
need not be Hermitian the left and right eigenstates may be different.
After undergoing the transformation in (\ref{eq:transform}) the 
disorder average of equation (\ref{eq:replica}) in
terms of $E_m$ and $\Psi_m$ is~\cite{bohr,hayn,hayn2}
\begin{multline}
\left\langle G^A(E,x,x)^{k-j}G^R(E,x,x)^j\right\rangle
=\lim_{n,n'\ra 0}
\frac{i^k(-1)^j\Gamma(n)\Gamma(n')}{\Gamma(k-j+n)\Gamma(j+n')}\\
\times\int db^{\dag}dbdb^{\prime\dag}db'\left(\sum_{l=1}^n
(b_{l\uparrow}^{\dag}b_{l\uparrow}
+b_{l\downarrow}b_{l\downarrow}^{\dag})
\right)^{k-j}\left(\sum_{l=1}^{n'}(b_{l\uparrow}^{\prime\dag}
b'_{l\uparrow}+b'_{l\downarrow}b_{l\downarrow}^{\prime\dag})
\right)^j\\
\times \sum_m|\Psi_m(b,b^{\dag},b',b^{\prime\dag})|^2
e^{-2LE_m}.\label{eq:replicag}
\end{multline}
The partition function for $n,n'\ra 0$ is defined as
\begin{equation}
Z_0=\lim_{n,n'\ra 0}\int\mathcal{D}s^*\mathcal{D}s\mathcal{D}
s^{\prime *}\mathcal{D}s' e^{-S},
\end{equation}
and when deriving equation (\ref{eq:replica}) we showed that $Z_0=1$.
After transforming as shown in equation (\ref{eq:transform})
the disorder averaged partition function in terms of $E_m$ and 
$\Psi_m$ is
\begin{equation}
\langle Z_0\rangle=\lim_{n,n'\ra 0}\int db^{\dag}dbdb^{\prime\dag}
db'\sum_m|\Psi_m(b,b^{\dag},b',b^{\prime\dag})|^2e^{-2LE_m}.
\end{equation}

As the system becomes infinite in length, that is 
$L\ra\infty$, the exponential term, $e^{-2LE_m}$ will vanish for all 
positive energy levels. In the the disorder averaged partition
function the dominant term is the ground state energy term
\begin{equation}
\langle Z_0\rangle=\lim_{n,n'\ra 0}\int db^{\dag}dbdb^{\prime\dag}
db'|\Psi_0(b,b^{\dag},b',b^{\prime\dag})|^2e^{-2LE_0}
\end{equation}
which may alternatively be written as
\begin{equation}
\langle Z_0\rangle=\lim_{n,n'\ra 0}\,_L\langle\Psi|e^{-2LE_0}
|\Psi\rangle_R
\end{equation}
where $|\Psi\rangle_L$ and $|\Psi\rangle_R$ are the left and right 
ground state wave functions respectively with
ground state energy $E_0$. As stated above
this average partition function must be unity so $E_0=0$ and 
\begin{equation}
\lim_{n,n'\ra 0}\,_L\langle\Psi|\Psi\rangle_R=1.\label{eq:normal}
\end{equation}
Hence, to calculate the Green's function correlator in the limit 
of an infinite
system we need only calculate the ground state of the transfer 
Hamiltonian, $H$. After taking the $L\ra\infty$ limit 
of (\ref{eq:replicag})
and using equation (\ref{eq:Zb}) to write it in terms of 
${\bf Z}$ and ${\bf Z}'$
\begin{multline}
\langle G^A(E,x,x)^{k-j}G^R(E,x,x)^j\rangle=\lim_{n,n'\ra 0}
\frac{(2i)^k(-1)^j\Gamma(n)\Gamma(n')}{\Gamma(k-j+n)\Gamma(j+n')}\\
\times\,_L\langle\Psi|Z_z^{k-j}Z_z^{\prime j}|\Psi\rangle_R.
\label{eq:replicaG}
\end{multline}
From equation (\ref{eq:mom}) the $k$th moment of the local density of
states, ${\tilde \rho}_1$, is
\begin{equation}
\langle{\tilde \rho}_1(\eta,E)^k\rangle=\lim_{n,n'\ra 0}\frac{k\mathrm{!}
\Gamma(n)\Gamma(n')}
{\pi^k}\sum_{j=0}^k\frac{_L\langle\Psi|Z_z^{k-j}Z_z^{\prime j}
|\Psi\rangle_R}{j\mathrm{!}
(k-j)\mathrm{!}\Gamma(k-j+n)\Gamma(j+n')}.
\end{equation}

\subsection{Supersymmetry}

We substitute the Dirac Hamiltonian (\ref{eq:h}) into equation 
(\ref{eq:susy}) to get the supersymmetric action
\begin{multline}
S=-\int_{-L}^L dy\left[\xi^*\sigma_z\partial\xi+s^*\sigma_z\partial 
s-\xi^{\prime *}\sigma_z\partial\xi'-s^{\prime *}\sigma_z\partial 
s'\right.\\
+i\Phi(\xi^*\sigma_x\xi+s^*\sigma_xs-\xi^{\prime *}\sigma_x\xi'-
s^{\prime *}\sigma_xs')+iM(\xi^*\sigma_y\xi+s^*\sigma_ys-
\xi^{\prime *}\sigma_y\xi'-s^{\prime *}\sigma_ys')\\
\left.+i(V-(E-i\eta))(\xi^*\xi+s^*s)-i(V-(E+i\eta))
(\xi^{\prime *}\xi'+s^{\prime *}s')\right].
\end{multline}
If we define the coordinates as
\begin{alignat}{2}
q_1&=s=(s_{\uparrow},s_{\downarrow}),& 
\qquad q_2&=\xi=(\xi_{\uparrow},\xi_{\downarrow}),\nonumber\\
q_1'&=s'=(s'_{\uparrow},s'_{\downarrow}),& \qquad q_2'&=\xi'=
(\xi'_{\uparrow},\xi'_{\downarrow}),
\end{alignat}
the momenta are
\begin{alignat}{2}
p_1&=(-s_{\uparrow}^*,s_{\downarrow}^*),& \qquad p_2&=
(-\xi_{\uparrow}^*,\xi_{\downarrow}^*),\nonumber\\
p_1'&=(s_{\uparrow}^{\prime *},-s_{\downarrow}^{\prime *}),& 
\qquad p_2'&=(\xi_{\uparrow}^{\prime *},-\xi_{\downarrow}^{\prime *}).
\end{alignat}

As in the replica case we need to transform this action. We shall use
the same transformation as Bocquet~\cite{bocquet} rather than the 
transformation used by Balents and Fisher~\cite{balents}
\begin{alignat}{2}
s_{\uparrow}&\ra b_{\uparrow},& \qquad \xi_{\uparrow}&\ra f_{\uparrow}
,\nonumber\\
s_{\downarrow}&\ra b_{\downarrow}^{\dag},& \qquad \xi_{\downarrow}&
\ra f_{\downarrow}^{\dag},\nonumber\\
s^*_{\uparrow}&\ra b^{\dag}_{\uparrow},& \qquad \xi^*_{\uparrow}&
\ra f^{\dag}_{\uparrow},\nonumber\\
s^*_{\downarrow}&\ra b_{\downarrow},& \qquad \xi^*_{\downarrow}&
\ra -f_{\downarrow}.
\end{alignat}
The negative sign in the transformation of the Grassman variables is
because for any Grassman variable $(\xi^*)^*=-\xi$. In terms of the
new variables the coordinates and momenta are
\begin{alignat}{2}
q_1&=(b_{\uparrow},b_{\downarrow}^{\dag}),& \qquad q_2&=(f_{\uparrow}
,f_{\downarrow}^{\dag}),\nonumber\\
q_1'&=(b'_{\uparrow},b_{\downarrow}^{\prime\dag}),& \qquad q_2'&=
(f'_{\uparrow},f_{\downarrow}^{\prime\dag}),\nonumber\\
p_1&=(-b_{\uparrow}^{\dag},b_{\downarrow}),& \qquad p_2&=
-(f_{\uparrow}^{\dag},f_{\downarrow}),\nonumber\\
p_1'&=(b_{\uparrow}^{\prime\dag},-b'_{\downarrow}),& \qquad p_2'&=
(f_{\uparrow}^{\prime \dag},f'_{\downarrow}).
\end{alignat}
The transformed action is
\begin{multline}
S=-\int_{-L}^L dy\left[f_{\uparrow}^{\dag}\partial f_{\uparrow}+
f_{\downarrow}\partial f_{\downarrow}^{\dag}-f_{\uparrow}^{\prime\dag}
\partial f'_{\uparrow}-f'_{\downarrow}\partial 
f_{\downarrow}^{\prime\dag}\right.\\
+b_{\uparrow}^{\dag}\partial b_{\uparrow}-b_{\downarrow}
\partial b_{\downarrow}^{\dag}-b_{\uparrow}^{\prime\dag}
\partial b'_{\uparrow}+b'_{\downarrow}\partial 
b_{\downarrow}^{\prime\dag}\\
+i\Phi(f_{\uparrow}^{\dag}f_{\downarrow}^{\dag}+
f_{\uparrow}f_{\downarrow}-f_{\uparrow}^{\prime\dag}
f_{\downarrow}^{\prime\dag}-f'_{\uparrow}f'_{\downarrow}
+b_{\uparrow}^{\dag}b_{\downarrow}^{\dag}+b_{\uparrow}
b_{\downarrow}-b_{\uparrow}^{\prime\dag}b_{\downarrow}^{\prime\dag}
-b'_{\uparrow}b'_{\downarrow})\\
+M(f_{\uparrow}^{\dag}f_{\downarrow}^{\dag}-f_{\uparrow}f_{\downarrow}
-f_{\uparrow}^{\prime\dag}f_{\downarrow}^{\prime\dag}+
f'_{\uparrow}f'_{\downarrow}+b_{\uparrow}^{\dag}b_{\downarrow}^{\dag}
-b_{\uparrow}b_{\downarrow}-b_{\uparrow}^{\prime\dag}
b_{\downarrow}^{\prime\dag}+b'_{\uparrow}b'_{\downarrow})\\
+i(V-(E-i\eta))(f_{\uparrow}^{\dag}f_{\uparrow}+f_{\downarrow}^{\dag}
f_{\downarrow}+b_{\uparrow}^{\dag}b_{\uparrow}+b_{\downarrow}^{\dag}
b_{\downarrow})\\
\left.-i(V-(E+i\eta))(f_{\uparrow}^{\prime\dag}f'_{\uparrow}
+f_{\downarrow}^{\prime\dag}f'_{\downarrow}+b_{\uparrow}^{\prime\dag}
b'_{\uparrow}+b_{\downarrow}^{\prime\dag}b'_{\downarrow})\right].
\end{multline}

We take the average over the disorder as described in equation
(\ref{eq:ave}). Following Bocquet~\cite{bocquet} we define two
superspins with three components 
\begin{align}
\mathcal{J}&=-\half p_1(i\sigma^x,i\sigma^y,\sigma^z)q_1
-\half p_2(i\sigma^x,i\sigma^y,\sigma^z)q_2,\nonumber\\
\mathcal{J}'&=\half p_1'(i\sigma^x,i\sigma^y,\sigma^z)q_1'
+\half p_2'(i\sigma^x,i\sigma^y,\sigma^z)q_2'.
\end{align}
Both these superspins satisfy $su(1,1)$ algebra
like ${\bf Z}'$ and ${\bf Z}$ in equations (\ref{eq:comm}). 
We can write the transfer Hamiltonian in terms of components 
of these two superspins
\begin{multline}
H=-2i\phi(\mathcal{J}_y-\mathcal{J}'_y)+2im(\mathcal{J}_x
-\mathcal{J}'_x)-2i(v-E)(\mathcal{J}_z-\mathcal{J}'_z)+2
\eta(\mathcal{J}_z+\mathcal{J}_z')\\
+4D_{\phi}(\mathcal{J}_y-\mathcal{J}'_y)^2+4D_m(\mathcal{J}_x
-\mathcal{J}'_x)^2+4D_v(\mathcal{J}_z-\mathcal{J}'_z)^2.\label{eq:hamj}
\end{multline}
Note that the transfer Hamiltonian, like the transfer Hamiltonian of
the replica trick, is independent of the spatial coordinate $x$.
We can decompose the superspins, $\mathcal{J}={\bf J}+{\bf S}$,
 into fermionic, ${\bf S}$, and
bosonic, ${\bf J}$, parts
\begin{alignat}{2}
{\bf J}&=-\half p_1(i\sigma^x,i\sigma^y,\sigma^z)q_1,\qquad 
&{\bf J}'&=\half p_1'(i\sigma^x,i\sigma^y,\sigma^z)q_1',\nonumber\\
{\bf S}&=-\half p_2(i\sigma^x,i\sigma^y,\sigma^z)q_2,\qquad 
&{\bf S}'&=\half p_2'(i\sigma^x,i\sigma^y,\sigma^z)q_2'.
\end{alignat} 
The ${\bf J}$, ${\bf J}'$, ${\bf S}$ and ${\bf S}'$ all satisfy the
$su(1,1)$ algebra. In terms of $b$, $b'$, $f$ and $f'$ 
\begin{alignat}{2}
J_x&=\textstyle{\frac{i}{2}}(b_{\uparrow}^{\dag}b_{\downarrow}^{\dag}
-b_{\downarrow}b_{\uparrow}),&\qquad
J'_x&=\textstyle{\frac{i}{2}}
(b_{\uparrow}^{\prime\dag}b_{\downarrow}^{\prime\dag}-
b'_{\downarrow}b'_{\uparrow}),\nonumber\\
J_y&=\half(b_{\uparrow}^{\dag}b_{\downarrow}^{\dag}
+b_{\downarrow}b_{\uparrow}),&\qquad J'_y&=
\half (b_{\uparrow}^{\prime\dag}
b_{\downarrow}^{\prime\dag}+b'_{\downarrow}b'_{\uparrow}),
\nonumber\\
J_z&=\half(b_{\uparrow}^{\dag}b_{\uparrow}
+b_{\downarrow}b_{\downarrow}^{\dag}),&\qquad J'_z&=\half 
(b_{\uparrow}^{\prime\dag}b'_{\uparrow}
+b'_{\downarrow}b_{\downarrow}^{\prime\dag}),\nonumber\\
S_x&=\textstyle{\frac{i}{2}}(f_{\uparrow}^{\dag}f_{\downarrow}^{\dag}
-f_{\downarrow}f_{\uparrow}),&\qquad
S'_x&=\textstyle{\frac{i}{2}}
(f_{\uparrow}^{\prime\dag}f_{\downarrow}^{\prime\dag}-
f'_{\downarrow}f'_{\uparrow}),\nonumber\\
S_y&=\half(f_{\uparrow}^{\dag}f_{\downarrow}^{\dag}
+f_{\downarrow}f_{\uparrow}),&\qquad S'_y&=
\half (f_{\uparrow}^{\prime\dag}
f_{\downarrow}^{\prime\dag}+f'_{\downarrow}f'_{\uparrow}),
\nonumber\\
S_z&=\half(f_{\uparrow}^{\dag}f_{\uparrow}
+f_{\downarrow}f_{\downarrow}^{\dag}),&\qquad S'_z&=\half 
(f_{\uparrow}^{\prime\dag}f'_{\uparrow}
+f'_{\downarrow}f_{\downarrow}^{\prime\dag}).\label{eq:J,S}
\end{alignat}

We can construct an integral like equation (\ref{eq:replicag}). The
supersymmetric version of this equation has, in
addition to the integrals over the complex or bosonic variables, $b$,
$b'$, $b^{\dag}$ and $b^{\prime\dag}$, 
integrals over the Grassman or fermionic variables, $f$, $f'$,
$f^{\dag}$ and $f^{\prime\dag}$. The $m$th eigenstate of the transfer
Hamiltonian, $\Psi_m$, has a dependence on the fermionic variables as
well as the bosonic variables. Like the replica trick we can define a 
disorder averaged partition function which is set to unity.
When we take the limit $L\ra\infty$ we obtain the supersymmetric
equivalent of equation (\ref{eq:replicaG})
\begin{equation}
\langle G^A(E,x,x)^{k-j}G^R(E,x,x)^j\rangle=
\frac{(2i)^k(-1)^j}{(k-j)\mathrm{!}j\mathrm{!}}\,_L\langle\Psi|
J_z^{k-j}J_z^{\prime j}|\Psi\rangle_R,
\label{eq:susyG}
\end{equation}
where, like the replica trick, $|\Psi\rangle_R$ and $|\Psi\rangle_L$
are the right and left ground 
state wave functions of the transfer
Hamiltonian with ground state energy $E_0=0$. 
From equation (\ref{eq:mom}) the $k$th moment of the local
 density of states, ${\tilde \rho}_1$, is
\begin{equation}
\langle{\tilde\rho}_1(\eta,E)^k\rangle=\frac{k\mathrm{!}}
{\pi^k}\sum_{j=0}^k\frac{_L\langle\Psi|J_z^{k-j}J_z^{\prime j}
|\Psi\rangle_R}{(j\mathrm{!}(k-j)\mathrm{!})^2}.\label{eq:denJ}
\end{equation}

The first moment of this local density of states is
\begin{equation}
\langle{\tilde\rho}_1(\eta,E)\rangle=\frac{1}
{\pi}\,_L\langle\Psi|J_z+J_z'|\Psi\rangle_R.\label{eq:denJ1}
\end{equation}
If we take the limit $\eta\ra 0$ this equation should be equivalent to 
equation (\ref{eq:density2}). 
Using equation (\ref{eq:J,S}) we can relate
$\psi_{\uparrow}$ and $\psi_{\downarrow}$ to the bosonic 
variables $b$ and $b'$.
To find the moments of the full local density 
of state, $\rho_1$, rather than ${\tilde \rho}_1$, 
 we must include the rapidly oscillating
terms in (\ref{eq:rapid}) . This means that instead of $J_z$ in 
equation (\ref{eq:denJ}) we have $J_z+\half J_+e^{-2ik_Fx}+\half
J_-e^{2ik_Fx}$, and similarly for $J_z'$. So,
\begin{multline}
\langle{\rho}_1(\eta,E)^k\rangle=\frac{k\mathrm{!}}
{\pi^k}\sum_{j=0}^k\frac{1}{(j\mathrm{!}(k-j)\mathrm{!})^2}
\,_L\langle\Psi|(J_z+\half J_+e^{-2ik_Fx}+\half
J_-e^{2ik_Fx})^{k-j}\\
\times(J_z'+\half J_+'e^{-2ik_Fx}+\half
J_-'e^{2ik_Fx})^j|\Psi\rangle_R.\label{eq:denJtot}
\end{multline}
Note that when this is expanded out terms involving $J_+J_-'$ and
$J_-J_+'$ can produce contributions to the moment that are slowly
varying in space. This argument to find $\langle\rho_1^k\rangle$
is also valid for the replica
trick. We used supersymmetry rather than the replica trick
because the notation is simpler.

\section{The ground state}
\label{sec-ground}
\subsection{Replica trick}
Bocquet~\cite{bocquet} has calculated the ground state for the 
transfer Hamiltonian associated with the one-particle retarded Green's
function. This ground state is a linear combination of the basis
states $\{|p\rangle\}$
\begin{equation}
|\Psi\rangle_R=\sum_{p=0}^{\infty}\zeta_{p}|p\rangle.
\end{equation}
This basis set is infinite due to the non-compactness of $u(1,1)$. 
The basis states are generated by a raising operator $Z_+$
\begin{equation}
|p\rangle=\frac{(Z_+)^k}{k\mathrm{!}}|\Omega\rangle,
\end{equation}
where $|\Omega\rangle$ is the lowest basis state. 
The basis states are orthogonal with norm
\begin{equation}
\langle p|p\rangle=\frac{\Gamma(p+n)}{\Gamma(n)\Gamma(p+1)}.
\end{equation}
As $n\ra 0$, $\langle p|p\rangle=\delta_{0p}$.
Bocquet's ${\bf Z}$ is identical to the ${\bf Z}'$ in this paper.

We use the same general form for the ground state of our transfer
Hamiltonian but summed over an additional variable 
\begin{equation}
|\Psi\rangle_R=\sum_{p, p'=0}^{\infty}\zeta_{pp'}|p,p'\rangle.
\end{equation}
The non-primed terms are related to the advanced Green's function and 
the primed terms are related to the retarded Green's function. The 
primed term should act in the same way as Bocquet's variables and the 
non-primed should act like complex conjugates.

We construct some raising and lowering operators, $Z_{\pm}=Z_y\mp 
iZ_x$ and $Z_{\pm}'=Z_y'\mp iZ_x'$ (as was done in section
\ref{sec-trans}). The basis states, $|p,p'\rangle$, 
are generated by the action of $Z_+'$
\begin{equation}
|p,p'\rangle=\frac{(Z_+')^{p'}}{p'\mathrm{!}}|p,\Omega'\rangle,
\end{equation}
and, since the non-primed terms act like complex conjugates of the
primed terms, the action of $Z_-$
\begin{equation}
|p,p'\rangle=\frac{(Z_-)^{p}}{p\mathrm{!}}|\Omega,p'\rangle.
\end{equation}
The state $|\Omega,\Omega'\rangle$ is the lowest vector in this 
representation. Alternatively, we could have $Z_+$ generating the
basis states since $(Z_-)^{\dag}=Z_+$
\begin{equation}
\langle p,p'|=\langle \Omega,p'|\frac{(Z_+)^{p}}{p\mathrm{!}}.
\end{equation}
 
The vector ${\bf Z}'$ operates on the $|p,p'\rangle$ basis state 
in the following way
\begin{align}
Z_z'|p,p'\rangle&=\left(p'+\textstyle{\frac{n'}{2}}\right)|p,p'\rangle
\nonumber\\
Z_+'|p,p'\rangle&=(p'+1)|p,p'+1\rangle
\nonumber\\
Z_-'|p,p'\rangle&=(p'+n'-1)|p,p'-1\rangle,
\label{eq:Z'}
\end{align}
and for ${\bf Z}$ operating on $\langle p,p'|$
\begin{align}
\langle p,p'|Z_z&=\left(p+\textstyle{\frac{n}{2}}\right)\langle p,p'|
\nonumber\\
\langle p,p'|Z_+&=(p+1)\langle p+1,p'|
\nonumber\\
\langle p,p'|Z_-&=(p+n-1)\langle p-1,p'|.
\label{eq:Z}
\end{align}

The scalar products of these states can be calculated by using, for
example, 
\begin{align}
(Z_-'|p,p'\rangle)^{\dag}&=\langle p,p'|Z_+'\nonumber\\
&=\langle p,p'-1|(p'+n'-1).
\end{align}
The norm is
\begin{align}
\langle p,p'|p,p'\rangle &=\left\langle p,p'\left|\frac{1}
{p\mathrm{!}p'\mathrm{!}}(Z_-)^p(Z_+')^{p'}\right|\Omega,\Omega'
\right\rangle\nonumber\\
&=\frac{\Gamma(p+n)\Gamma(p'+n')}{\Gamma(n)\Gamma(n')\Gamma(p+1)
\Gamma(p'+1)}\langle\Omega,\Omega'|\Omega,\Omega'\rangle\nonumber\\
&=\frac{\Gamma(p+n)\Gamma(p'+n')}{\Gamma(n)\Gamma(n')\Gamma(p+1)
\Gamma(p'+1)}.\label{eq:pp'}
\end{align}
Note that ${\bf Z}$ and ${\bf Z}'$ commute. In the limit $n,n'\ra 0$
\begin{equation}
\langle p,p'|p,p'\rangle =\delta_{0p}\delta_{0p'}.
\end{equation}
Similarly, it can also be shown that the scalar product of
 $|p,p'\rangle$ and $|q,q'\rangle$ with 
$p\neq q$ or $p'\neq q'$ vanishes, assuming $\langle
p,p'|\Omega,\Omega'\rangle=0$ if $p\neq\Omega$ and $p'\neq\Omega'$. 
Hence, the basis states are orthogonal. 

As discussed previously, because the transfer Hamiltonian isn't 
Hermitian the left wave function is not the Hermitian conjugate 
of the right wave function. The Hamiltonian and its Hermitian 
conjugate are related by the transform $R=\exp(i\pi(Z_z+Z_z'))$
~\cite{bocquet}. The left and right 
ground state wave functions are related by this same transform
\begin{equation}
|\Psi\rangle_L=R|\Psi\rangle_R.
\end{equation}
Using equation (\ref{eq:pp'}) and the orthogonality of the basis
states 
\begin{equation}
_L\langle\Psi|\Psi\rangle_R=\frac{1}{\Gamma(n)\Gamma(n')}
\sum_{p,p'=0}^{\infty}(-1)^{p+p'+\frac{n}{2}+\frac{n'}{2}}
\frac{\Gamma(p+n)\Gamma(p'+n')}{\Gamma(p+1)\Gamma(p'+1)}
|\zeta_{pp'}|^2.\label{eq:psi}
\end{equation}
As $n,n'\ra 0$ this reduces to
\begin{equation}
_L\langle\Psi|\Psi\rangle_R=|\zeta_{00}|^2
\end{equation}
so we can make our normalization condition $|\zeta_{00}|^2=1$, in
agreement with equation (\ref{eq:normal}).

Using equations (\ref{eq:Z'}), (\ref{eq:Z}) and (\ref{eq:psi}) 
\begin{multline}
_L\langle\Psi|Z_z^{k-j}Z_z^{\prime j}|\Psi\rangle_R=
\frac{1}{\Gamma(n)\Gamma(n')}\sum_{p,p'=0}^{\infty}(-1)^{p+p'
+\frac{n}{2}+\frac{n'}{2}}\left(p+\frac{n}{2}\right)^{k-j}\\
\times\left(p'+\frac{n'}{2}\right)^j
\frac{\Gamma(p+n)\Gamma(p'+n')}
{\Gamma(p+1)\Gamma(p'+1)}|\zeta_{pp'}|^2.\label{eq:psiZ}
\end{multline}
Thus, by substitution into equation (\ref{eq:replicaG}),
\begin{multline}
\langle G^A(E,x,x)^{k-j}G^R(E,x,x)^j\rangle=
\frac{(2i)^k(-1)^j}{\Gamma(k-j)\Gamma(j)}
\sum_{p,p'=0}^{\infty}(-1)^{p+p'}\\
\times p^{k-j-1}p^{\prime j-1}
|\zeta_{pp'}|^2.
\label{eq:replicaG2}
\end{multline}
If we can find the ground state of the transfer Hamiltonian we can 
calculate all the $\zeta_{pp'}$ and hence we have calculated the 
Green's function correlator. We will calculate 
$\zeta_{pp'}$ for a special 
case of the Dirac equation corresponding to a single channel quantum
wire in section \ref{sec-calc}.

\subsection{Supersymmetry}

The bosonic parts of the superspins, ${\bf J}$ and ${\bf J}'$, are 
defined in exactly the same way as the ${\bf Z}$ and ${\bf Z}'$ 
used in the 
replica trick calculation when $n=n'=1$. So, ${\bf J}$ and ${\bf J}'$ 
satisfy all the same equations as ${\bf Z}$ and ${\bf Z}'$ when we set 
$n=n'=1$ in all the replica trick calculations. 
Like the replica trick we have a bosonic wave function 
$|p,p'\rangle$ but here we will use the 
notation $|p\rangle|p'\rangle'$. This bosonic wave function is not 
affected by the fermionic operators ${\bf S}$ and ${\bf S}'$. From 
equations (\ref{eq:Z'}) and (\ref{eq:Z})
\begin{alignat}{2}
J_z'|p\rangle'&=\left(p+\half \right)|p\rangle', &\qquad \langle p|J_z 
&=\left(p+\half \right)\langle p|,\nonumber\\
J_+'|p\rangle'&=(p+1)|p+1\rangle', &\qquad \langle p|J_+&=(p+1)
\langle p+1|,\nonumber\\
J_-'|p\rangle'&=p|p-1\rangle', &\qquad \langle p|J_-&=p\langle p-1|
,\label{eq:J}
\end{alignat}
where $J_{\pm}=J_y\mp iJ_x$ and $J_{\pm}'=J_y'\mp iJ_x'$.
From equation (\ref{eq:pp'})
\begin{equation}
\langle p|p\rangle=\,'\langle p|p\rangle'=1.\label{eq:Jpp'}
\end{equation}
The fermionic operators, ${\bf S}$ and ${\bf S}'$, are like spin 
operators. There are two spin up states and two spin down states 
and the fermionic operators operate on them as follows
\begin{alignat}{2}
S'_z|\uparrow\rangle'&=\half |\uparrow\rangle', &\qquad \langle
\uparrow|S_z&=\half \langle\uparrow|,\nonumber\\
S_+'|\uparrow\rangle'&=0, &\qquad \langle\uparrow|S_+&=0, 
\nonumber\\
S_-'|\uparrow\rangle'&=-|\downarrow\rangle', &\qquad \langle
\uparrow|S_-&=-\langle\downarrow|,\nonumber\\
S'_z|\downarrow\rangle'&=-\half |\downarrow\rangle', 
&\qquad \langle\downarrow|S_z&=-\half \langle\downarrow|,
\nonumber\\
S_+'|\downarrow\rangle'&=|\uparrow\rangle', &\qquad \langle
\downarrow|S_+&=\langle\uparrow|,\nonumber\\
S_-'|\downarrow\rangle'&=0, &\qquad \langle\downarrow|S_-&=0,
\label{eq:S}
\end{alignat}
where $S_{\pm}=S_y\mp iS_x$ and $S_{\pm}'=S_y'\mp iS_x'$. 
Note that the operation of $S_-$ and 
$S_-'$ on the spin up has the opposite sign from usual. This is due 
to the Grassman variables which make up the spin operators. For the 
same reason it can be shown that $(S_z)^3=(S_z')^3=0$. 

Balents and Fisher~\cite{balents} show that the basis state for 
the non-primed bosons and fermions is 
\begin{equation}
|p\rangle_R=\left\{\begin{array}{ll}
|p\rangle|\downarrow\rangle+|p-1\rangle|\uparrow\rangle,\qquad 
& p>0,\\
|0\rangle|\downarrow\rangle,\qquad & p=0,
\end{array}\right.
\end{equation}
and similarly for $|p\rangle_R'$. The total basis state combines
primed and non-primed parts and is written as 
$|p\rangle_R|p\rangle_R'$. The left basis state for $|p\rangle_R$ 
is
\begin{align}
|p\rangle_L&=(-1)^{\mathcal{J}_z}|p\rangle_R\nonumber\\
&=\left\{\begin{array}{ll}
(-1)^p(|p\rangle|\downarrow\rangle+|p-1\rangle|\uparrow\rangle),
\qquad & p>0,\\
|0\rangle|\downarrow\rangle, & p=0,
\end{array}\right.\label{eq:left}
\end{align}
and similarly for $|p\rangle_L'$.

As discussed in Bocquet~\cite{bocquet} the norm of the lowest basis
state, $|0\rangle_R$, can be taken to be unity
\begin{equation}
_L\langle 0|0\rangle_R=\langle 0|0\rangle\langle\downarrow|
\downarrow\rangle=1.
\end{equation} 
Since the $|p\rangle$'s are orthonormal the norm for the
down spins must be unity. The up spin can be defined in terms of the
down spin:
\begin{equation}
\langle\uparrow|=\langle\downarrow|S_+.
\end{equation}
This allows us to write the norm of the up spin as
\begin{equation}
\langle\uparrow|\uparrow\rangle=\langle\downarrow|S_+
S_-|\uparrow\rangle,
\end{equation}
which can be evaluated using the spin rules of equation
(\ref{eq:S}). We can calculate the norms of the primed spins in
the same way. The norms for all the spin states are 
\begin{equation}
-'\langle\uparrow|\uparrow\rangle'=\,'\langle\downarrow|\downarrow
\rangle'=-\langle\uparrow|\uparrow\rangle=\langle\downarrow|
\downarrow\rangle=1.\label{eq:spins}
\end{equation}

The total left and right 
wave functions are created by summing over all the basis states
\begin{align}
|\Psi\rangle_L&=\sum_{p,p'=0}^{\infty}\zeta_{pp'}|p\rangle_L
|p'\rangle_L',\nonumber\\
|\Psi\rangle_R&=\sum_{p,p'=0}^{\infty}\zeta_{pp'}|p\rangle_R
|p'\rangle_R',\label{eq:susypsi}.
\end{align}
Using the rules given above and the normalization condition on the
wave function it can be shown that
\begin{equation}
_L\langle\Psi|\Psi\rangle_R=|\zeta_{00}|^2=1.
\end{equation}

Combining equations (\ref{eq:J}) and (\ref{eq:S}) we obtain 
the following equations for $\mathcal{J}$ and $\mathcal{J}'$
\begin{alignat}{2}
\mathcal{J}_z'|p\rangle_R'&=p|p\rangle_R',
\qquad &_L\langle p|\mathcal{J}_z&=
\, _L\langle p|p,\nonumber\\
\mathcal{J}_+'|p\rangle_R'&=(p+1)|p+1\rangle_R',
\qquad & _L\langle p|\mathcal{J}_+&=
\, _L\langle p+1|(p+1),\nonumber\\
\mathcal{J}_-'|p\rangle_R'&=(p-1)|p-1\rangle_R',
\qquad & _L\langle p|\mathcal{J}_-&=
\, _L\langle p-1|(p-1)\label{eq:calJ}.
\end{alignat}
These are exactly the same as ${\bf Z}$ and ${\bf Z}'$ for the replica 
trick for $n=n'=0$. 

To calculate (\ref{eq:susyG}) we require
\begin{equation}
_L\langle\Psi|J_z^{k-j}J_z^{\prime j}|\Psi\rangle_R=\,_L
\langle\Psi|(\mathcal{J}_z-S_z)^{k-j}(\mathcal{J}'_z-S'_z)^j
|\Psi\rangle_R.\label{eq:psiJ}
\end{equation}
This expression can be expanded using equations (\ref{eq:S}) and 
(\ref{eq:calJ}) and by using the form of the ground state in 
equation (\ref{eq:susypsi}) 
\begin{align}
J_z^{k-j}J_z^{\prime j}|\Psi\rangle_R&=(\mathcal{J}_z-S_z)^{k-j}
(\mathcal{J}'_z-S'_z)^j|\Psi\rangle_R\nonumber\\
&=\left(\mathcal{J}_z^{k-j}-(k-j)\mathcal{J}_z^{k-j-1}S_z+\half 
(k-j)(k-j-1)\mathcal{J}_z^{k-j-2}S_z^{2}\right)\nonumber\\
&\times\left(\mathcal{J}_z^{\prime j}-j\mathcal{J}_z^{\prime j-1}
S_z'+\half j(j-1)\mathcal{J}_z^{\prime j-2}S_z^{\prime 2}\right)
|\Psi\rangle_R\nonumber\\
&=\sum_{p,p'=0}^{\infty}\left(p^{k-j}-(k-j)p^{k-j-1}S_z+\half 
(k-j)(k-j-1)p^{k-j-2}S_z^{2}\right)\nonumber\\
&\times\left(p^{\prime j}-jp^{\prime j-1}S_z'
+\half j(j-1)p^{\prime j-2}S_z^{\prime 2}\right)
\zeta_{pp'}|p\rangle_R|p'\rangle'_R\nonumber\\
&=\sum_{p,p'=0}^{\infty}\left[\left(p^{\prime j}+
\half jp^{\prime j-1}
+\textstyle{\frac{1}{8}}j(j-1)p^{\prime j-2}\right)
|p'\rangle'|\downarrow\rangle'\right.\nonumber\\
&\left.+\left(p^{\prime j}-\half jp^{\prime j-1}+
\textstyle{\frac{1}{8}}j(j-1)p^{\prime j-2}\right)
|p'-1\rangle'|\uparrow\rangle'\right]\nonumber\\
&\times\left[\left(p^{k-j}+\half (k-j)p^{k-j-1}+
\textstyle{\frac{1}{8}}(k-j)(k-j-1)p^{k-j-2}\right)\right.
|p\rangle|\downarrow\rangle\nonumber\\
&+\left(p^{k-j}-\half (k-j)p^{k-j-1}+\textstyle{
\frac{1}{8}}(k-j)(k-j-1)p^{k-j-2}\right)\nonumber\\
&\left.\times|p-1\rangle|
\uparrow\rangle\right]\zeta_{pp'},
\end{align}
where, in order to simplify the above formula, we have set
$|-1\rangle_R=|-1\rangle_R'=0$.
Using equations (\ref{eq:spins}) and (\ref{eq:left}) and the 
orthogonality of the basis states 
\begin{multline}
_L\langle\Psi|J_z^{k-j}J_z^{\prime j}|\Psi\rangle_R=
\sum_{p,p'=1}^{\infty}(-1)^{p+p'}|\zeta_{pp'}|^2(k-j)j
p^{k-j-1}p^{\prime j-1}\\
+\sum_{p=1}^{\infty}(-1)^p|\zeta_{p0}|^2(k-j)p^{k-j-1}
\left(p^{\prime j}+\half j
p^{\prime j-1}+\textstyle{\frac{1}{8}}j(j-1)p^{\prime j-2}
\right)_{p'=0}\\
+\sum_{p'=1}^{\infty}(-1)^{p'}|\zeta_{0p'}|^2jp^{\prime j-1}
\left(p^{k-j}+\half (k-j)
p^{k-j-1}\right.\\
\left.+\textstyle{\frac{1}{8}}(k-j)(k-j-1)p^{k-j-2}
\right)_{p=0}\\
+\left(p^{k-j}+\half (k-j)p^{k-j-1}+
\textstyle{\frac{1}{8}}(k-j)(k-j-1)p^{k-j-2}\right)_{p=0}\\
\times\left(p^{\prime j}+\half jp^{\prime j-1}+
\textstyle{\frac{1}{8}}j(j-1)p^{\prime j-2}\right)_{p'=0}.
\label{eq:susyG2}
\end{multline}
The terms arising from $p=0$ will vanish for $k-j>2$. The terms
due to $p'=0$ will vanish for $j>2$. 
The above equation may be
substituted into equation (\ref{eq:susyG}). When $k-j>2$ and $j>2$ 
this equation will resemble equation (\ref{eq:replicaG2}) but 
with the $Z_z$ and $Z'_z$ replaced with $J_z$ and $J'_z$. 
For $k-j\leq 2$ or $j\leq 2$ the supersymmetric form
will have some additional constant terms. However, as will be shown 
in the following section, the density of states calculated from 
supersymmetry does reduce to the density of states calculated from 
the replica trick. As for replica trick we
need to solve for $\zeta_{pp'}$ in order to calculate the Green's
function correlator.

\section{Evaluation of the moments of the local density of states}
\label{sec-randommass}
We shall now evaluate the moments of the local density of states 
for a specific case corresponding to a disordered quantum wire with a
single channel. We set
$\phi=m=v=D_{v}=0$ and $D_{\phi}=D_{m}=D$.

\subsection{Solution of the ground state wave function of the 
transfer Hamiltonian}
\label{sec-calc}

The replica trick transfer Hamiltonian (\ref{eq:ham}) with 
$\phi=m=v=D_{v}=0$ and $D_{\phi}=D_{m}=D$ is
\begin{multline}  
H=2iE(Z_z-Z_z')+2\eta(Z_z+Z_z')\\
+2D\left((Z_+-Z_+')(Z_--Z_-')+(Z_--Z_-')(Z_+-Z_+')\right).
\end{multline}
The supersymmetric transfer Hamiltonian (\ref{eq:hamj}) has a 
similar form but with the vectors ${\bf Z}$ and ${\bf Z}'$ 
replaced with the superspins $\mathcal{J}$ and $\mathcal{J}'$. 
Also, as can be seen by comparing equations 
(\ref{eq:Z'}) and (\ref{eq:Z}) with equation (\ref{eq:calJ}) 
the operation of ${\bf Z}$ on the replica ground state with $n=n'=0$ 
is identical to the operation of the $\mathcal{J}$ on the
supersymmetric ground state. Hence, the ground state eigenvalue
equations must be 
identical when we take $n$ and $n'$ to zeroth order in the replica 
ground state. We shall calculate the zeroth order replica ground state 
here and use the same ground state in the supersymmetry calculations.

If substitute the transfer Hamiltonian into
\begin{equation}
\langle p,p'|H|\Psi\rangle=0,
\end{equation}
using, for example, 
\begin{align}
\langle p,p'|\Psi\rangle=&\zeta_{pp'}\nonumber\\
\langle p,p'|Z_z|\Psi\rangle=&p\zeta_{pp'}\nonumber\\
\langle p,p'|Z_+Z_-|\Psi\rangle=&(p+1)p\zeta_{pp'}\nonumber\\
\langle p,p'|Z_+Z_-'|\Psi\rangle=&pp'\zeta_{p+1p'+1},
\end{align}
we obtain
\begin{multline}
iE(p-p')\zeta_{pp'}+\eta(p+p')
\zeta_{pp'}\\
+2D(p^2\zeta_{pp'}+p^{\prime 2}\zeta_{pp'}-pp'\zeta_{p-1p'-1}
-pp'\zeta_{p+1p'+1})=0.
\end{multline}
This equation implies that $\zeta_{pp'}^*=\zeta_{p'p}$ which implies
that $\zeta_{pp}$ is real. 
If we assume that $\zeta_{pp'}=\delta_{pp'}\zeta_{pp}$ then
\begin{equation}
\eta p\zeta_{pp}+Dp^2(2\zeta_{pp}-\zeta_{p-1p-1}
-\zeta_{p+1p+1})=0.
\end{equation}
Note that this equation is the same as equation (47) in 
Reference~\cite{altshuler} which was obtained using the 
Berezinskii technique.  
We define a new variable $x=\eta p/D$. As $p$ increases to
$p+1$ $x$ only increases by $\eta/D$. It is reasonable to assume that
$\eta\ll D$ so we can take $x$ to be a continuous variable. If
$\zeta_{pp}=\zeta(x)$ we obtain the following
differential equation of $\zeta$ in terms of $x$
\begin{equation}
\frac{d^2\zeta(x)}{dx^2}-\frac{1}{x}\zeta(x)=0,
\end{equation}
which has the general solution
\begin{equation}
\zeta(x)=x^{1/2}\left[a'K_1(2x^{1/2})
+bI_1(2x^{1/2})\right],
\end{equation}
where $a'$ and $b'$ are constants. In terms of $p$
\begin{equation}
\zeta_{pp}=p^{1/2}\left[aK_1\left(2\sqrt{\frac{\eta p}{D}}\right)
+bI_1\left(2\sqrt{\frac{\eta p}{D}}\right)\right],
\end{equation}
where $a$ and $b$ are constants.

Boundary conditions require that $\zeta$ remains finite for all $p$ 
so we must have $b=0$. We can find the other constant of 
integration, $a$, by the normalization condition on the ground state. 
The normalizing condition is $|\zeta_{00}|^2=1$.
We also know that $\zeta_{00}$ must be real so
\begin{equation}
\zeta_{00}=1=a\lim_{p\ra 0}p^{\half }K_1\left(2\sqrt{\frac{\eta p}
{D}}\right),
\end{equation}
which gives
\begin{equation}
a=2\sqrt{\frac{\eta}{D}}.
\end{equation}
Hence
\begin{equation}
|\Psi\rangle_R=\sum_{p=0}^{\infty}2\sqrt{\frac{\eta m}{D}
}K_1\left(2\sqrt{\frac{\eta m}{D}}\right)|p,p\rangle
\end{equation}
and
\begin{equation}
\zeta_{pp'}=\delta_{pp'}2\sqrt{\frac{\eta p}{D}}K_1\left(2
\sqrt{\frac{\eta p}{D}}\right).\label{eq:zeta}
\end{equation}

\subsection{Replica trick}

By substituting equation (\ref{eq:zeta})
 into (\ref{eq:replicaG2})
\begin{align}
\langle G^A(E,x,x)^{k-j}G^R(E,x,x)^j\rangle&=\frac{(2i)^k(-1)^j}
{\Gamma(k-j)\Gamma(j)}\frac{4\eta}{D}
\sum_{p=0}^{\infty}p^{k-1} K_1\left(2\sqrt{
\frac{p\eta}{D}}\right)^2\nonumber\\
&=\frac{(2i)^k(-1)^j}{\Gamma(k-j)\Gamma(j)}
\left(\frac{D}{\eta}\right)^{k-1}\frac{k}{k-1}
\frac{\Gamma(k)^4}{\Gamma(2k)}.\label{eq:g}
\end{align}
We have used the fact that in the limit $\eta\ra 0$ the 
sum over $p$ can be converted into an integral by defining the
variable $x=2\sqrt{\eta p/D}$ which may be approximated to
a continuous variable since $\eta\ll D$
\begin{align}
\frac{4\eta}{D}\sum_{p=0}^{\infty}p^{k-1}K_1
\left(2\sqrt{\frac{\eta p}{D}}\right)^2&=
2\left(\frac{D}{4\eta}\right)^{k-1}\int_0^{\infty}dxx^{2k-1}
K_1(x)^2\nonumber\\
&=\left(\frac{D}{\eta}\right)^{k-1}\frac{k}{k-1}\frac{
\Gamma(k)^4}{\Gamma(2k)},\label{eq:sum-int}
\end{align}
and~\cite{ryzhik}
\begin{equation}
\int_0^{\infty}x^{\lambda}K_1(x)^2dx=2^{\lambda-2}
\frac{\lambda+1}{\lambda-1}\frac{\Gamma\left(
\frac{\lambda+1}{2}\right)^4}{\Gamma(\lambda+1)},
\end{equation}
if $\lambda>1$. Note that as $\eta\ra 0$ the sum in equation
(\ref{eq:g}) is dominated by the terms with large $p$. Hence, the
non-compactness of $u(1,1)$ is playing an essential role here.  

By substituting (\ref{eq:g}) into equation (\ref{eq:mom}) we 
obtain the moments of the density of states
\begin{equation}
\langle{\tilde \rho}_1^k\rangle=\frac{1}{\pi^k}
\left(\frac{D}{\eta}\right)^{k-1}
\frac{\Gamma(k+1)}{(2k-1)}\label{eq:d}
\end{equation}
by using the relation (calculated by Maple 5.1)
\begin{equation}
k\mathrm{!}\sum_{j=0}^k\frac{(k-j)j}{((k-j)\mathrm{!}j
\mathrm{!})^2}=\frac{(k-1)\Gamma(2k-1)}{\Gamma(k)^3}.\label{eq:jsum}
\end{equation}
When $k=1$ we find that $\langle{\tilde \rho}_1\rangle=1/\pi$.
This equation for $\langle{\tilde\rho}_1^k\rangle$ agrees with equation 
(69) in Al'tshuler and Prigodin~\cite{altshuler} and, more 
recently~\cite{feldmann}, where 
$s_1=8\eta\tau=\eta/D$. The two results 
differ by a factor of $1/\pi^k$ because what Al'tshuler and Prigodin
define as $\langle\rho^k\rangle$ we define as 
$\langle\rho^k\rangle/\langle\rho\rangle^k$, and similarly for all the
density of states. 

\subsection{Supersymmetry}

To calculate the moments of the density of states from supersymmetry
we substitute equation (\ref{eq:zeta}) into equation 
(\ref{eq:susyG2}). The terms from $p=0$ and $p'=0$ will be negligible
for small $\eta$ for all cases except $k=1$. From equation
(\ref{eq:susyG2}) and for $k>1$ we obtain the same result as for 
the replica trick, shown in equation (\ref{eq:g}). The 
moments of the density of states are given in equation (\ref{eq:d}).

The calculations for $k=1$ are straight forward. It is easy to show
\begin{equation}
_L\langle\Psi |J_z|\Psi\rangle_R=\,_L\langle\Psi| J_z'|\Psi\rangle_R
=\half 
\end{equation}
which gives $\langle{\tilde \rho}_1\rangle=\frac{1}{\pi}$ which 
agrees with equation (\ref{eq:d}).   

\section{Moments of the full density of states}
\label{sec-fulldos}
We showed at the end of section \ref{sec-trans} how to calculate
the moments of the full density of states, $\langle\rho_1^k\rangle$,
described by equation (\ref{eq:denJtot}). We require these moments 
 rather than the moments of the density of states which removes 
the rapidly varying terms, $\langle{\tilde \rho}_1^k\rangle$, 
described by equation 
(\ref{eq:denJ}). By comparing these two equations we see that to
obtain the moments of the full density of states instead of
$J_z^{\prime j}$ we need the term 
\begin{equation}
\sum_{m=0}^j\sum_{l=0}^m(\half)^{j-m+l}J_+^{\prime j-m}J_-^{\prime l}
J_z^{\prime m-l}\frac{j\mathrm{!}e^{2ik_F(l+m-j)}}
{(j-m)\mathrm{!}(m-l)\mathrm{!}l\mathrm{!}}.\label{eq:sum}
\end{equation}
A similar expression may be written for the additional non-primed 
terms. We have ignored that fact that the components of ${\bf J}'$ 
do not commute since, as will be shown, the terms arising from 
interchanging the components are negligible. 
Instead of $J_z^{k-j}J_z^{\prime j}$ we use
\begin{multline}
\sum_{m=0}^j\sum_{r=0}^{k-j}\sum_{l=0}^m\sum_{q=0}^r(\half)^{k-m-r+l+q}
J_+^{\prime j-m}J_-^{\prime l}J_z^{\prime m-l}
J_+^{k-j-r}J_-^qJ_z^{r-q}\\
\times\frac{j\mathrm{!}(k-j)\mathrm{!}e^{2ik_F(m+r+l+q-k)}}
{(j-m)\mathrm{!}(m-l)\mathrm{!}l\mathrm{!}(k-j-r)
\mathrm{!}(r-q)\mathrm{!}q\mathrm{!}}.
\end{multline}
Now we can neglect those terms which contain rapidly varying
exponentials. We retain
those terms for which $q=k-l-m-r$. However $q$ must be in the range
$[0,r]$. This places a restriction on $l$ 
\begin{equation}
\textrm{max}(0,k-m-2r)\leq l\leq\textrm(m,k-m-r).
\end{equation}
The above sum is now
\begin{multline}
\sum_{m=0}^j\sum_{r=0}^{k-j}
\sum_{l=\textrm{max}(0,k-m-2r)}^{\min(m,k-m-r)}
(\half)^{2(k-m-r)}
J_+^{\prime j-m}J_-^{\prime l}J_z^{\prime m-l}
J_+^{k-j-r}J_-^{k-m-r-l}J_z^{2r+m+l-k}\\
\times\frac{j\mathrm{!}(k-j)\mathrm{!}}
{(j-m)\mathrm{!}(m-l)\mathrm{!}l\mathrm{!}(k-j-r)
\mathrm{!}(2r+m+l-k)\mathrm{!}(k-m-r-l)\mathrm{!}}.
\end{multline}

We need to calculate something of the form
$_L\langle\Psi|J_+^{\prime a}J_-^{\prime b}J_z^{\prime c}
J_+^dJ_-^eJ_z^f|\Psi\rangle_R.$
To get insight into how this is done we first evaluate
\begin{multline}
_L\langle\Psi|J_+^{\prime a}J_-^{\prime b}
J_z^{\prime c}|\Psi\rangle_R
=\sum_{p=0}^{\infty}|\zeta_{pp}|^2(p-1)\ldots(p-b+1)(p-b+1)
\ldots(p-b+a-1)\\
\times[p(p-b+a)(p^c+\half cp^{c-1}+\textstyle{\frac{1}{8}}
(c-1)(c-2)p^{c-2})\\
-(p-b)^2(p^c-\half cp^{c-1}+\textstyle{\frac{1}{8}}
(c-1)(c-2)p^{c-2})],
\end{multline}
where we have used $\zeta_{pp'}=\delta_{pp'}\zeta_{pp}$.
As $\eta\ra 0$ the sum is dominated by terms with large $p$ (since
$\zeta_{pp}$ is proportional to $\eta$) and so  we
retain just those terms with the highest power of $p$ which is
$p^{a+b+c-1}$. This is why in equation (\ref{eq:sum}) we can swap the
components of ${\bf J}'$ around, because the extra terms are a smaller
power of $p$ than $p^{a+b+c-1}$ which become negligible as $\eta\ra
0$. The simplified result is
\begin{equation}
_L\langle\Psi|J_+^{\prime a}J_-^{\prime b}
J_z^{\prime c}|\Psi\rangle_R
=\sum_{p=0}^{\infty}|\zeta_{pp}|^2(a+b+c)p^{a+b+c-1}.
\end{equation}
When we include both the primed and non-primed terms we obtain
\begin{equation}
_L\langle\Psi|J_+^{\prime a}J_-^{\prime b}
J_z^{\prime c}J_+^dJ_-^eJ_z^f|\Psi\rangle_R
=\sum_{p=0}^{\infty}|\zeta_{pp}|^2(a+b+c)(d+e+f)p^{a+b+c+d+e+f-2}.
\end{equation}

Substituting these identities into equation (\ref{eq:denJtot}) gives
\begin{multline}
\langle\rho_1(\eta,E)^k\rangle=\frac{k\mathrm{!}}{\pi^k}
\sum_{p=0}^{\infty}
|\zeta_{pp}|^2p^{k-2}\sum_{j=0}^k\frac{j(k-j)}{j\mathrm{!}
(k-j)\mathrm{!}}\sum_{m=0}^j\sum_{r=0}^{k-j}\\
\times\sum_{l=\textrm{max}(0,k-m-2r)}^{\min(m,k-m-r)}
(\half)^{2(k-m-r)}\\
\times[(j-m)\mathrm{!}(m-l)\mathrm{!}l\mathrm{!}(k-j-r)
\mathrm{!}(2r+m+l-k)\mathrm{!}(k-m-r-l)\mathrm{!}]^{-1}.
\end{multline}
The sum over $p$ can be evaluated as in the previous section using
equation (\ref{eq:zeta}) and (\ref{eq:sum-int}). 

The sum over $j$, $m$, $r$ and $l$ can be rewritten 
as~\cite{efetov2}
\begin{multline}
S=\sum_{j=0}^k\frac{j(k-j)}{j\mathrm{!}
(k-j)\mathrm{!}}\sum_{m=0}^j\sum_{r=0}^{k-j}
\sum_{l=0}^m\sum_{l'=k-m-2r}^{k-m-r}
\delta_{ll'}(\half)^{2(k-m-r)}\\
\times[(j-m)\mathrm{!}(m-l)\mathrm{!}l\mathrm{!}(k-j-r)
\mathrm{!}(2r+m+l'-k)\mathrm{!}(k-m-r-l')\mathrm{!}]^{-1}.
\end{multline}
If we write the kronecker delta as
\begin{equation}
\delta_{ll'}=\frac{1}{2\pi}\int_0^{2\pi}e^{i(l-l')t}dt
\end{equation}
and rewrite the sum over $l'$ as a sum over $q=l'-k-m+2r$ it 
can be shown that
\begin{equation}
S=\frac{(k-1)\Gamma(2k-1)}{k\Gamma(k)^4}\frac{1}{2\pi}\int_0
^{2\pi}dt(\textstyle{\frac{1}{4}}e^{-it}+1+e^{it})^k,
\end{equation}
by making use of equation (\ref{eq:jsum}).
The integral can be evaluated after a binomial expansion. After
integration the resulting sum may be simplified using Maple 5.1 so
that
\begin{equation}
S=\frac{(k-1)(2k-1)\Gamma(2k-1)^2}{2^{k-1}k^2\Gamma(k)^6}.
\end{equation} 

The moments of the total density of states is thus
\begin{equation}
\langle\rho_1(\eta,E)^k\rangle=\frac{1}{\pi^k}\left(\frac{D}{2\eta}
\right)^{k-1}\frac{\Gamma(2k-1)}{\Gamma(k)}.
\label{eq:mom2}
\end{equation}
This agrees with equation (68) in Al'tshuler and
Prigodin~\cite{altshuler} although we have an additional factor of
$1/\pi^k$, as discussed previously,
due to different scaling of the density of states.
Using equation (\ref{eq:mom-pratio}) we can use the moments of the 
density of states in equation (\ref{eq:mom2})
to calculate the $k$th moment of the participation ratio
\begin{equation}
P^{(k)}=(2D)^{k-1}\Gamma(k),\label{eq:partratio}
\end{equation}
where we have used $\langle\rho\rangle=
\lim_{\eta\ra 0}\langle\rho_1\rangle=1/\pi$. 
 
\section{Probability distribution function}
\label{sec-dist}
The moments of the participation ratio may be used to find
moments of density of states, $\langle\rho_f^k\rangle$,
 which are derived from some arbitrary
weight $f$ as defined in equation (\ref{eq:rho_f}). 
The probability distribution function of any local density of states,
$W(\rho_f)$, may be obtained from these moments of the density of     
states by the following~\cite{altshuler}
\begin{equation}
\int_{0}^{\infty}W(\rho)\exp(-p\rho)d\rho=\exp\left[\sum_{k=1}
^{\infty}\frac{(-p)^k\langle\rho^k_f\rangle_c}
{k\mathrm{!}\langle\rho_f\rangle^k}\right].
\label{eq:dist}
\end{equation}
The cumulants are defined by 
\begin{equation}
\langle\rho_f^k\rangle_c=\langle\rho_f^k\rangle-
\langle\rho_f^{k-1}\rangle.
\end{equation} 
Solving equation (\ref{eq:dist}) for $W$ requires an inverse 
Laplace transform.

Following Al'tshuler and Prigodin~\cite{altshuler} we now summarise
how one can construct a density of states which is related to the NMR
line shape. In a metal the NMR line undergoes a frequency shift, known
as the Knight shift, which
is proportional to the local density of states derived using
(\ref{eq:gend}) with weight
\begin{equation}
f_2(E)=-\frac{n_F(E+g\mu_eH)-n_F(E-g\mu_eH)}{2g\mu_eH}.
\end{equation} 
This weight is the derivative of the Fermi distribution function, 
$n_F$, about the Zeeman splitting $g\mu_eH$. 
The Fermi distribution function is
\begin{equation}
n_F(E)=\left[\exp\left(\frac{E-\epsilon_F}{T}\right)+1\right]^{-1},
\end{equation}
where $\epsilon_F$ is the Fermi energy. 
The density of states derived from this weight within the region 
$T\gg g\mu_eH$ will be defined as $\rho_2(T,\epsilon_F)$. In this high
temperature limit
\begin{equation}
f_2(E)=\frac{1}{4T}\cosh^{-2}\left(\frac{E-\epsilon_F}{2T}\right).
\label{eq:f2}
\end{equation}
The local density of states, $\rho_2$, fluctuates throughout
the sample. Hence, the Knight shift also fluctuates. The net result of
these various Knight shifts is a broadening of the NMR line shape. If
the relaxation rate is of the sample is less than the Knight shift the
NMR line shape is fully determined by the superposition of all the
local Knight shifts. This is proportional to the distribution of the
density of states, $\rho_2$.

\begin{figure}[b]
  \begin{center}
  \epsfxsize=\hsize
  \epsfbox{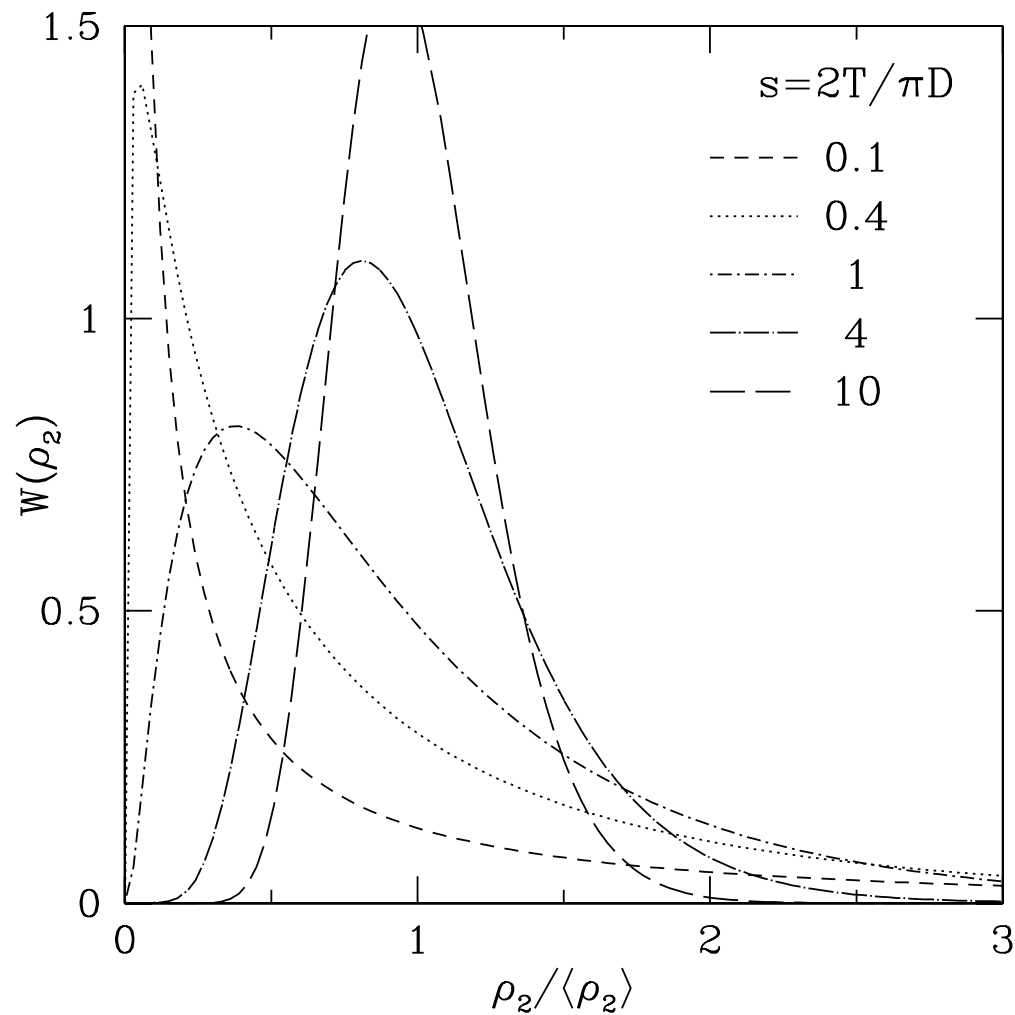}
\caption[The probability distribution function
for a number of different values of $s$.]{The probability distribution 
function, equation (\ref{eq:w}),
for a number of different values of $s=2T/\pi D$. For large $s$ (high
temperature and low disorder) the distribution is symmetric so that
the typical value of the density of states is the same as the average
value. As $s$ decreases (low temperature and high disorder) the
distribution becomes skewed and there is a marked difference between
the typical and the average value of the density of states. 
This probability distribution can be related to the NMR line shape.
}
  \label{fig:w}
  \end{center}
\end{figure}

By substituting equation (\ref{eq:partratio}) and (\ref{eq:f2}) into
equation (\ref{eq:rho_f}) we find the moments of this new density of 
states are
\begin{equation}
\langle\rho_2(T,\epsilon_F)^k\rangle=\left(\frac{D}{2T}\right)^{k-1}
\frac{\Gamma(k)^2\Gamma(\textstyle{\frac{3}{2}})}{\pi\Gamma(k+\half)},
\label{eq:momrho2}
\end{equation}
where we have used~\cite{ryzhik}
\begin{equation}
\int_{-\infty}^{\infty}dy\cosh^{-2k}y=2\frac{\Gamma(k)
\Gamma(\textstyle{\frac{3}{2}})}{\Gamma(k+\half)}.
\end{equation}
The sum in equation (\ref{eq:dist}) is dominated by the large $k$
terms. If $k$ is large we can assume that the $k$th cumulant is 
approximately equal to the $k$th moment of the density of states
\begin{equation}
\langle\rho_2^k\rangle_c=\langle\rho_2^k\rangle. 
\end{equation}
After substituting the moments shown in equation (\ref{eq:momrho2})
into equation (\ref{eq:dist}) and taking the inverse Laplace transform
Al'tshuler and Prigodin~\cite{altshuler} obtained
\begin{equation}
W(\rho_2)=\frac{s}{2\pi}\int_{-\infty}^{\infty}dx\sinh{2x}
\sin(\pi sx)
\exp\left(-s\left[\frac{\rho_2}{\langle\rho_2\rangle}
\cosh^2{x}+x^2-\frac{\pi}{4}\right]\right),\label{eq:w}
\end{equation}
where $s=2T/\pi D$ in our notation but $s=16T\tau/\pi$ in Al'tshuler and
Prigodin's notation. The relationship between the disorder strength,
$D$, and the scattering time, $\tau$, is $8\tau=D^{-1}$. 

The probability distribution function at large temperature 
of the density of states, $\rho_2$, is plotted in figure 
\ref{fig:w} for a number of values of $s$. Note that for $s\leq 1$ the
distribution is quite asymmetric and the typical (most probable) value
is significantly less than the average value.  
We have set various constants to unity, such as the Boltzmann
constant, $k_B$, the Fermi velocity, $v_F$, and $\hbar$. If all these
constants are included we find that near the Fermi energy
$s=2k_BT\hbar v_F/\pi D$ or, in terms of the mean free path,
$l=v_F\tau=(\hbar v_F)^2/8D$, we have $s=16k_BTl/\pi\hbar 
v_F=6.67\times 10^{11}\,Tl/v_F$. If we have a sample with a mean 
free path of $1\,\mu$m and a
Fermi velocity of $v_F=1\times 10^6$\,ms$^{-1}$ then  
some skewness in the NMR line shape should be observable when the 
temperature is 1.5\,K or less ($s\leq 1$). 

\section{Conclusion}

We have presented a general formalism
to calculate exactly the disorder average of a product of 
Green's functions, $\langle G^A(E,x,x)^{k-j}G^R(E,x,x)^j\rangle$, 
for the random Dirac equation using both the replica
trick and supersymmetry.
This extends the formalism developed
recently to calculate the disorder average of
one Green's function using 
the replica trick~\cite{bocquet} and 
supersymmetry~\cite{balents,bocquet}.
The problem is reduced to that of finding
the ground state of a zero-dimensional Hamiltonian
involving a pair of ``spins'' that are elements of $u(1,1)$.
The moments of the different density of states can be written
in terms of the expectation values of various ``spin'' operators
in this ground state.
We explicitly found this ground state wave function 
and the associated expectation values for the case
of the Dirac equation corresponding 
to a disordered quantum wire with a single channel.  
The non-compactness of $u(1,1)$ played an essential
role in the results.
We showed how the moments of the local density 
of states changed when it was averaged over
atomic length scales.
Our results for the moments agree
with those obtained previously by
Al'tshuler and Prigodin~\cite{altshuler} using 
the Berezinskii diagram method.
In our view, our derivation is more transparent from
a field-theoretic point of view.
Future work should focus on finding the ground 
state of the transfer Hamiltonian corresponding
to more general random Dirac equations, particularly
those associated with quantum phase transitions
in random spin chains.

\ack
We thank I. A. Gruzberg, J. B. Marston and M. Gould 
for helpful discussions. 
This work was supported by the Australian Research Council.

\appendix
\section{APPENDIX}
\label{app-int}
In this appendix we derive the prefactor in the replica trick
expansion of the $k$th power of the Green's function shown in 
equation (\ref{eq:coeff}).
Consider an $N$ dimensional integral of the form 
\begin{equation}
I_N=\int d^Nx F(x_l^2)
\end{equation}
where $F(x_l^2)$ is a function which depends only on the length 
$r=\sum_{l=1}^Nx_l^2$. By making a change of variables into polar 
coordinates:
\begin{equation}
(x_1,x_2,\ldots,x_N)\ra(r,\phi,\theta_1,\ldots,\theta_{N-2})
\end{equation}
we can evaluate the integrals over the $\theta$'s and the $\phi$ 
(since $F(x_l^2)=F(r)$ has no dependence on the $\theta$'s or the 
$\phi$) and we are left with~\cite{ramond}
\begin{equation}
I_N=\frac{\pi^{N/2}}{\Gamma(N/2)}\int_0^{\infty}dr r^{\frac{N-2}{2}}
F(r)\label{eq:in}.
\end{equation}

For example, if $s$ and $s^*$ are one dimensional variables such that 
$s=x_1+ix_2$ and $A$ is a constant
\begin{equation}
\frac{1}{A^k}=\frac{1}{k\mathrm{!}}\left[\int ds^*ds(s^*s)^k
e^{-s^*As}\right]\left[\int ds^*ds e^{-s^*As}\right]^{-1}
\label{eq:1/Ak}
\end{equation}
then the integral has dimension $N=2$ and $F(r)=r^ke^{-Ar}$ where 
$r=s^*s=x_1^2+x_2^2$. Using equation (\ref{eq:in}) it can be shown 
that in terms of $r$ the integral is
\begin{equation}
\frac{1}{A^k}=\frac{1}{k\mathrm{!}}\left[\int_0^{\infty} 
drr^ke^{-Ar}\right]
\left[\int_0^{\infty} dr e^{-Ar}\right]^{-1}.
\end{equation}
 
We now consider a replicated version of equation (\ref{eq:1/Ak}). If
this equation is replicated $n$ times
\begin{multline}
\frac{1}{A^k}=\frac{X}{k\mathrm{!}}\left[\int \prod_{l=1}^nds_l^*
ds_l\left(\sum_{l=1}^ns_l^*s_l\right)^k\exp\left(-\sum_{l=1}^n
s_l^*As_l\right)\right]\\
\times\left[\int \prod_{l=1}^nds_l^*ds_l \exp\left(-\sum_{l=1}^n
s_l^*As_l\right)\right]^{-1}
\end{multline}
where $X$ is a factor which must be calculated. We can write 
$s_l=x_{2l-1}+ix_{2l}$ so now we are dealing with a $N=2n$ 
dimensional integral and $r=\sum_{l=1}^ns_l^*s_l=\sum_{l=1}^{2n}
x_l^2$. The integral becomes
\begin{equation}
\frac{1}{A^k}=\frac{X}{k\mathrm{!}}\frac{\pi^n/\Gamma(n)}{\pi^n/
\Gamma(n)}\left[\int_0^{\infty} drr^{n-1}r^ke^{-Ar}\right]\left[
\int_0^{\infty} dr
r^{n-1}r^{-Ar}\right]^{-1}.
\end{equation}
The integrals over the $r$ variable may be evaluated to show that
\begin{equation}
\frac{1}{A^k}=\frac{X}{k\mathrm{!}}\frac{1}{A^k}
\frac{\Gamma(n+k)}{\Gamma(n)}
\end{equation}
so
\begin{equation}
X=\frac{k\mathrm{!}\Gamma(n)}{\Gamma(n+k)}
\end{equation}
and hence the replicated form of equation (\ref{eq:1/Ak}) is
\begin{multline}
\frac{1}{A^k}=\frac{\Gamma(n)}{\Gamma(n+k)}\left[\int 
\prod_{l=1}^nds_l^*ds_l\left(\sum_{l=1}^ns^*s\right)^k
\exp\left(-\sum_{l=1}^ns_l^*As_l\right)\right]\\
\times\left[\int \prod_{l=1}^nds_l^*ds_l \exp\left(-\sum_{l=1}^n
s_l^*As_l\right)\right]^{-1}.
\end{multline}
Note that when $n=1$ we get back to equation (\ref{eq:1/Ak}). 
If $k=1$ then the factor before the integral is $1/n$ which 
agrees with Bocquet~\cite{bocquet}.



\listoffigures

\end{document}